\documentclass[draft]{agujournal2019}
\usepackage{url} 
\usepackage{lineno}
\usepackage{multirow}
\usepackage{rotating}
\usepackage{booktabs}

\usepackage{soul}

\newcommand{\apjs}{{\it Astrophys. J. Suppl. S.}}
\newcommand{\aap}{{\it Astron. Astrophys.}}

\draftfalse

\journalname{Space Weather}

\begin{document}

%
%


\title{First observations of a geomagnetic superstorm with a sub-L1 monitor}

%
%




\authors{E. Weiler\affil{1,2}, C. M\"ostl\affil{1}, E. E. Davies\affil{1}, A. M. Veronig\affil{2,3}, U. V. Amerstorfer\affil{1}, T. Amerstorfer\affil{1}, J. Le Lou\"edec\affil{1}, M. Bauer\affil{1,2}, N. Lugaz\affil{4}, V. Haberle\affil{5}, H. T. R\"udisser\affil{1,2}, S. Majumdar\affil{1}, M. Reiss\affil{6}}

\affiliation{1}{Austrian Space Weather Office, GeoSphere Austria, Reininghausstraße 3, Graz, 8020, Austria}
\affiliation{2}{Institute of Physics, University of Graz, Universit\"atsplatz 5, Graz, 8010, Austria}
\affiliation{3}{Kanzelh\"ohe Observatory for Solar and Environmental Research, University of Graz, Kanzelh\"ohe 19, Treffen am Ossiacher See, 9521, Austria}
\affiliation{4}{Space Science Center and Department of Physics and Astronomy, University of New Hampshire, 8 College Rd, Durham, NH 03824, USA}
\affiliation{5}{Conrad Observatory, GeoSphere Austria, Hohe Warte 38, Vienna, 1190, Austria}
\affiliation{6}{Community Coordinated Modeling Center, NASA Goddard Space Flight Center, 8800 Greenbelt Rd., Greenbelt, MD 20771, USA}




\correspondingauthor{Eva Weiler}{eva.weiler@geosphere.at}



\begin{keypoints}
\item The strongest geomagnetic storm since 2003 occurred on 10--12 May 2024 and provided a unique test case for future space weather missions.
\item We discuss the solar sources and interplanetary evolution of five interacting coronal mass ejections that caused this event.
\item Data from STEREO-A at a sub-L1 position extended the lead time and provided a fairly accurate prediction of the geomagnetic storm magnitude.
\end{keypoints}

%
%

%
%


\begin{abstract} 
Forecasting the geomagnetic effects of solar coronal mass ejections (CMEs) is currently an unsolved problem. CMEs, responsible for the largest values of the north-south component of the interplanetary magnetic field, are the key driver of intense and extreme geomagnetic activity. Observations of southward interplanetary magnetic fields are currently only accessible directly through in situ measurements by spacecraft in the solar wind. On 10--12 May 2024, the strongest geomagnetic storm since 2003 took place, caused by five interacting CMEs. We clarify the relationship between the CMEs, their solar source regions, and the resulting signatures at the Sun--Earth L1 point observed by the ACE spacecraft at 1.00~AU.
The STEREO-A spacecraft was situated at 0.956~AU and 12.6$^\circ$ west of Earth during the event, serving as a fortuitous sub-L1 monitor providing interplanetary magnetic field measurements of the solar wind. We demonstrate an extension of the prediction lead time, as the shock was observed 2.57~hours earlier at STEREO-A than at L1, consistent with the measured shock speed at L1, 710~km$\,\mathrm{s}^{-1}$, and the radial distance of 0.043~AU. By deriving the geomagnetic indices based on the STEREO-A beacon data, we show that the strength of the geomagnetic storm would have been decently forecasted, with the modeled minimum SYM-H$\,=-\,478.5$~nT, underestimating the observed minimum by only 8\%. Our study sets an unprecedented benchmark for future mission design using upstream monitoring for space weather prediction. 
\end{abstract}

\section*{Plain Language Summary}
Severe space weather, caused by geoeffective solar storms, has been recognized as a natural hazard. Studies have shown that extreme geomagnetic storms, occurring up to once every 50 to 100 years, may disrupt technological systems. However, predicting the effects of these solar storms is difficult. The necessary information about the storm's internal magnetic field near Earth can currently only be collected by spacecraft at a specific point in space called L1, which is 0.01~AU in front of Earth and provides only 10 to 60 minutes warning time. In May 2024, the strongest geomagnetic storm since 2003 occurred, caused by five interacting solar eruptions. This study examines the solar origin of these storms and their relation to the data collected by the ACE spacecraft at L1. We also use data from another spacecraft, STEREO-A, which was slightly closer to the Sun than L1 during the event. By analyzing the STEREO-A data, we show that the strength of the storm could have been predicted fairly accurately 2~hours and 34~minutes earlier than with data just from L1. Our study provides a new standard for planning future missions using spacecraft at distances closer to the Sun than L1 to predict space weather.

%
%

%


%
%
%
%

\section{Introduction}

Coronal mass ejections (CMEs) are huge bursts of plasma from the Sun and the main cause of severe space weather phenomena at Earth. However, not every CME directed towards the Earth is geoeffective \cite<e.g.>{kilpua2019review,temmer2023}. Whether a CME that hits the Earth causes a geomagnetic storm depends on its embedded magnetic field. Specifically, it matters whether the magnetic field component parallel to the Earth's magnetic axis is directed southward. In that case, magnetic reconnection with the oppositely directed Earth magnetic field can occur. Although we can draw conclusions about the structure and propagation of CMEs in the heliosphere from remote sensing, the prediction of the southward magnetic field component is still an unsolved problem in space weather research, known as the $B_z$ problem \cite<e.g.>[]{vourlidas2019review}.

The geomagnetic impact of CMEs is generally quantified via geomagnetic indices, one of the most widely used being the disturbance storm time ($Dst$) index, measured in nanotesla (nT) and provided by the World Data Center for Geomagnetism, Kyoto \cite{sugiura1964,kyoto_dst}. This measure has been available since 1957, and thus, is well suited for comparison with historic storms. The $Dst$ index quantifies the reduction of the horizontal magnetic field on the ground due to the enhancement of the magnetospheric ring current and is calculated every hour from four mid-latitude ground observatories. It is used to characterize the strength of geomagnetic storms, with minimum $Dst$ indices of less than -50, -100, -200, and -250~nT indicating moderate, severe, intense, and superstorms, respectively \cite<e.g.>[]{villaverde2023}. CMEs are the exclusive driver for intense events \cite<e.g.>[]{zhang2007}, with successive and possibly interacting CMEs thought to be the cause of the most severe storms \cite<e.g.>[]{liu2014observations,meng2019superstorms,koehn2022}. Geomagnetic superstorms can cause damage to satellites, power grids and disrupt communications, making the development of appropriate forecasting and mitigation measures a necessity for our technology-dependent society.

Currently, we rely mainly on in situ observations from spacecraft positioned at the first Lagrangian Point (L1), located 0.01~AU upstream (sunward) of Earth, to obtain reliable estimates of the severity of geomagnetic storms. Some of these spacecraft, namely the Advanced Composition Explorer \cite<ACE;>[]{stone1998ace} and the NOAA-operated Deep Space Climate Observatory \cite<DSCOVR;>[]{dscovr}, transmit interplanetary magnetic field and plasma data in real-time, with a delay of only a few minutes, allowing us to predict the geomagnetic effects of CMEs 10 to 60 minutes before the CME reaches Earth.

\citeA{burton1975} was the first to find an empirical relationship to forecast the ground-based $Dst$ index, and thus the strength of the geomagnetic storm, using the measured interplanetary magnetic field and plasma parameters. Since then, many others have worked on creating solar wind-to-$Dst$ index models, ranging from empirical and semi-empirical to machine learning modeling approaches \cite<e.g.>{obrien2000,lundstedt2002,boynton2011}. \citeA{ji2012} found that the \citeA{Temerin2006} model (henceforth referred to as the TL model) achieved the best prediction performance for severe geomagnetic events out of 63 different models. This model is semi-empirical and was optimized using eight years of solar wind data, achieving a root mean square error (RMSE) of 6.65~nT between the observed and predicted $Dst$ index.

On 10--12 May 2024, the strongest geomagnetic storm since 2003 took place, with a minimum $Dst$ of -412~nT, which is thought to be around 30--50\% the magnitude of the most extreme events \cite<e.g.>[]{riley2012frequency,love2015_historic, love2024carrington}. Consequentially, problems with the positioning global navigation satellite systems, the drop in orbital height of satellites, the rerouting of flights, and strong geoelectric fields have been reported, but no major damages nor unplanned outages of power grids were observed \cite<e.g.>{Hayakawa2024,Spogli2024,themens2024}. Mitigation measures for geomagnetically induced currents have been reported, for example, in the USA and New Zealand, with the relevant guidelines for New Zealand given in \citeA<>[]{manus2023NZmitigation}. On the bright side, for tens, if not hundreds of millions of people, the chain of events presented the first time to experience the northern and southern lights as the auroral oval expanded to lower latitudes. 

As the superstorm was measured not only by spacecraft at L1 (at 1.00~AU), but also by the Solar TErrestrial RElations Observatory \cite<STEREO-A;>{kaiser2008stereo} positioned at 0.96~AU and 12.6$^\circ$ west of Earth (located at 1.01~AU), this event gives us the first opportunity to study how a spacecraft situated closer to the Sun than L1 could improve both the lead time and accuracy for forecasting the geomagnetic effects of superstorms. This is particularly interesting with regard to future mission concepts for sub-L1 monitors, which are intended to improve the predictability of space weather by deploying spacecraft further upstream than L1 \cite<e.g.>{lindsay1999dst, stcyr2000diamond, kubicka2016dst, morley2020, laker_2024, lugaz2024mission}. 

Depending on the location of the sub-L1 monitor, the warning time for large-scale disturbances in the solar wind, including CMEs but also high-speed streams, could be extended by several hours. Various strategies have been proposed to achieve this goal, in particular the stationing of preferably multiple spacecraft on distant retrograde orbits \cite<DROs;>{henon1969,stcyr2000diamond,Borovsky2018}, on a Venus-like orbit \cite{ritter2015}, or the placement of a spacecraft around imaginary Lagrange points using solar sails \cite{lindsay1999dst, eastwood2015}. The spacecraft for such missions must naturally lie in the ecliptic plane and must not deviate too much from the Sun-Earth line in order to ensure a certain degree of accuracy, regarding the longitudinal coherence of CMEs and processes that influence CME propagation. The ideal orbital distance of these spacecraft has not yet been determined, as there is a lack of multi-spacecraft observations of CMEs at medium separations (1$^\circ$ to 20$^\circ$ in heliospheric longitude). However, recent studies \cite<e.g.>[]{good2016interplanetary,lugaz2024MEwidth} have shown that the longitudinal separation between two spacecraft should be less than 15$^\circ$ to ensure that most CME magnetic ejecta are actually measured at both spacecraft. Since STEREO-A  has a longitudinal separation of 12.6$^\circ$, the May 2024 event is considered to be in the upper longitudinal separation range for which monitoring at distances closer to the Sun than L1 is still feasible. As it is the first event that led to a geomagnetic superstorm being observed by a sub-L1 monitor, it provides a unique opportunity to test the real-time predictive capability of superstorms from a sub-L1 perspective. To this end, we perform a hindcasting analysis and model the geomagnetic effects from the STEREO-A data using only data and knowledge available at the time of the event.

In this study, we examine the origin of the CMEs that caused the May event and their propagation in remote images, connecting these to in situ observations at both STEREO-A and L1, and compare the predicted and observed geomagnetic effects. The first three sections give a chronological summary of the event. We start with an analysis of the source region and describe the remote sensing data used for our study and the derivations thereof in section~\ref{sec:solar_interplanetary_evolution}. In section~\ref{sec:elevo} we simulate the propagation of the CMEs in the heliosphere using the semi-empirical propagation tool ELEvo. The in situ magnetic structures of the CMEs as measured by STEREO-A and ACE are analyzed in section~\ref{sec:in_situ}, where we link those to their solar and interplanetary counterparts, as inferred from solar and heliospheric observations, respectively. We explain the TL model and the processing of the STEREO-A beacon data in section~\ref{sec:results}. The resulting modeled geomagnetic indices from real-time STEREO-A and L1 data as well as a comparison to observed indices are presented in the same section. Finally, in section~\ref{sec:conclusion} our results and main findings are summarized and discussed.

\section{Solar and Remote Sensing Observations}\label{sec:solar_interplanetary_evolution}

To study the source regions of these CMEs we use different filters of the Solar Dynamics Observatory \cite<SDO;>{pesnell2012sdo} Atmospheric Imaging Assembly \cite<AIA;>{lemen2011atmospheric} as well as the Helioseismic and Magnetic Imager \cite<HMI;>{scherrer2012HMI}. Furthermore, we use $\mathrm{H} \alpha$ observations from the Cerro Tololo observatory in Chile, which is one of six telescopes of the Global Oscillation Network Group \cite<GONG;>{harvey1996GONG}.

\begin{figure}[h!]
    \centering
    \includegraphics[width=\linewidth]
    {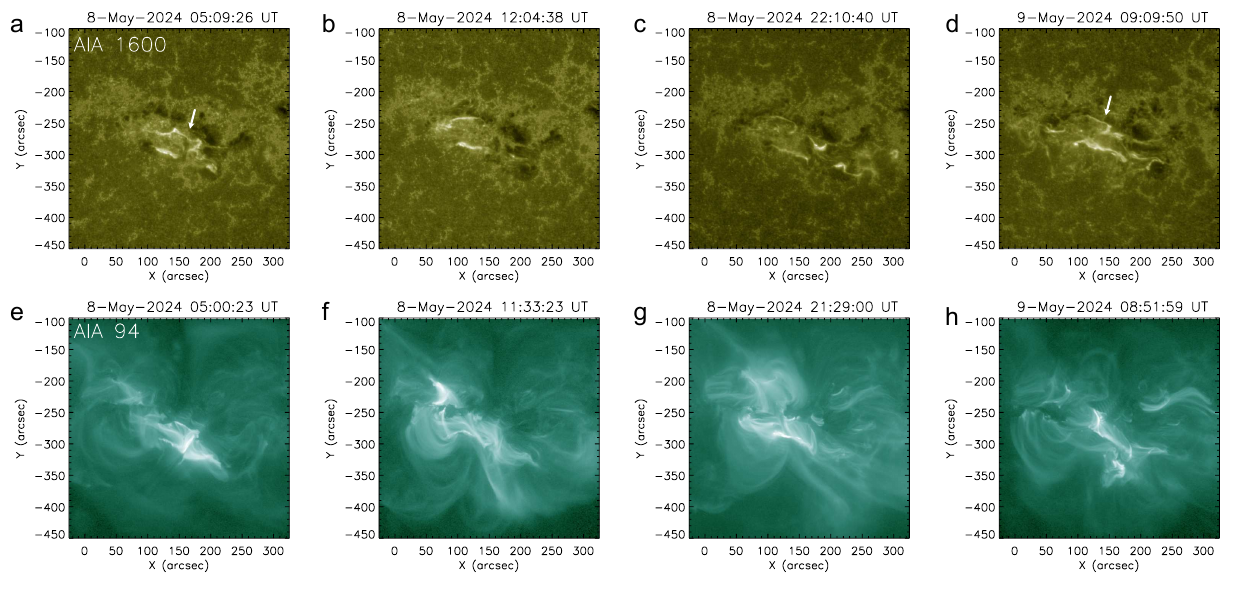}
    \caption{Snapshots of the four events from AR 13664 (associated with CME 1, CME 2, CME 4, and CME 5) from SDO/AIA 1600 {\AA} and 94 {\AA} filtergrams, (a)--(d) imaging the flare ribbons in the chromosphere and (e)--(h) the hot flaring corona (at a temperature of $\sim$~7~MK). All images are differentially rotated to 8 May 2024, 05:00~UT. The white arrows in (a) and (d) highlight forward J-shaped flare ribbons.}
    \label{fig:source_region_eruptions}
\end{figure}

\begin{figure}[h!]
    \centering
    \includegraphics[width=0.95\linewidth]
    {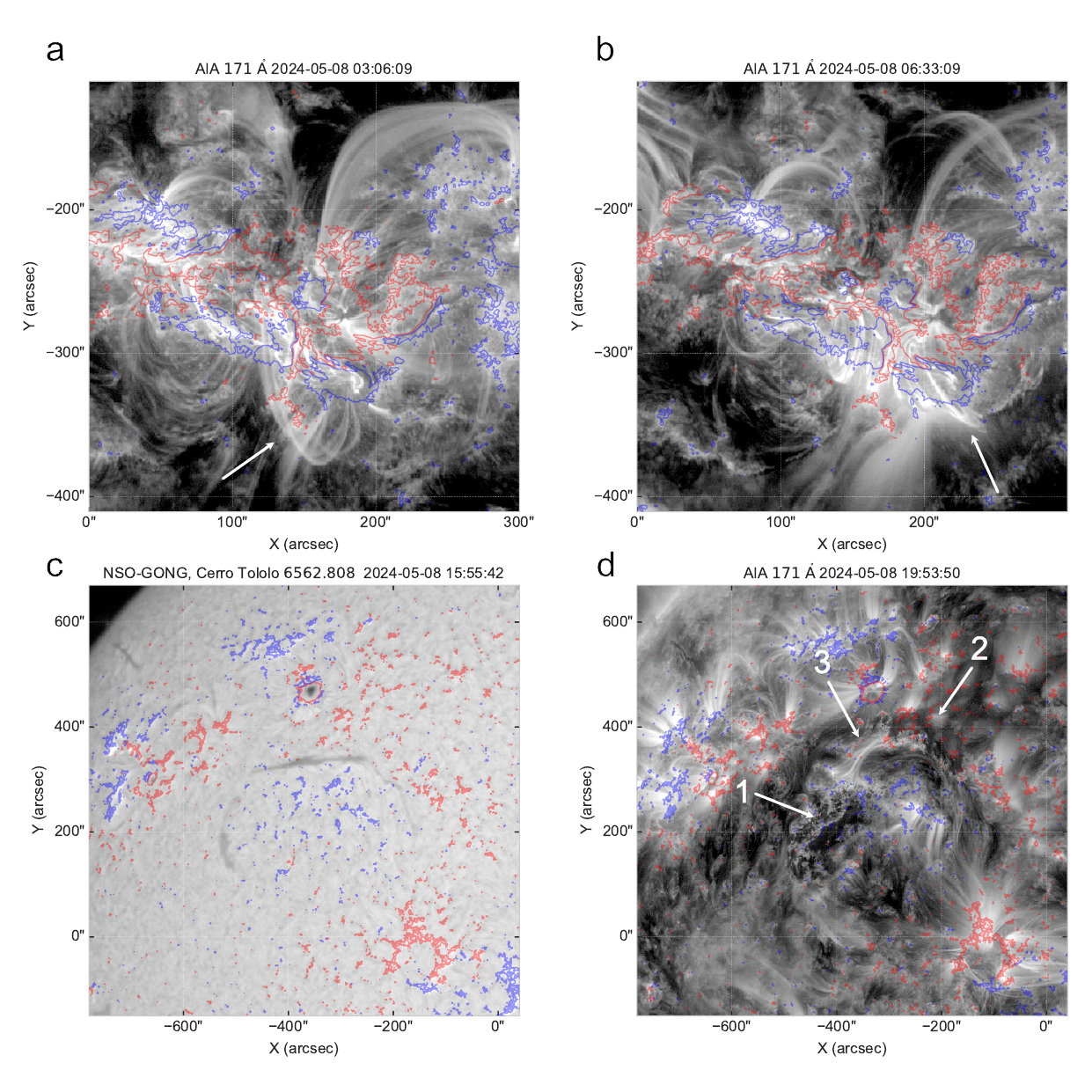}
    \caption{Solar proxies to derive magnetic helicity. (a)--(b) AR 13664 in the southern hemisphere observed in the AIA 171~{\AA} filter, with the arrow in (a) showing the overlying coronal loops, and the arrow in (b) showing the flare arcades. (c) Filament in northern hemisphere as observed in $\mathrm{H} \alpha$ by GONG, Cerro Tololo observatory. (d) Dimming region (arrows 1 and 2) and skew of post-eruptive arcades (arrow 3) as seen in the AIA 171~{\AA} filter. The snapshots are overlaid with HMI magnetograms saturated at $\pm$100~G (a)--(b) and $\pm$300~G (c)--(d). Red colors indicate positive polarity, blue colors negative polarity.}
    \label{fig:handedness}
\end{figure}

Figure~\ref{fig:source_region_eruptions} shows the active region AR 13664. Four of the five CMEs (numbers 1, 2, 4, and 5) that caused the main geomagnetic effect on 10--11 May 2024 originated from this solar active region and were associated with major flares of soft X-ray class X1.0, M8.3, X1.0/M9.8 (double peak, two-step eruption), and X2.2 (see Table \ref{tab:initial_parameters_flare}), whereas CME 3 originated from a different region on the Sun. During 8--9 May 2024, AR 13664 was located close to disk center (20$^\circ$ S, 10--20$^\circ$ W), was of complex magnetic type ($\beta\gamma\delta$), and covered a size as large as about 1200 micro-hemispheres. 
Starting from 7 May 2024, it showed a steep increase in its free magnetic energy, with a value of about $1\cdot 10^{33}$ erg on 8 May 2024 \cite{Jarolim2024,Hayakawa2024}.
The flares related to CMEs 1 and 2 occurred in the eastern (trailing) part of the elongated AR 13664; the flares associated with CMEs 4 and 5 also had their center in the AR's trailing segment, but in addition also revealed flare ribbons extending to its western segment. Each of these four CMEs were associated with impulsive hot loop eruptions observed by the 94 and 131~{\AA} filters (Figure~\ref{fig:source_region_eruptions}e--h), which have peak formation temperatures of $\sim$~7 and 10~MK, respectively. From the 94~{\AA} filter, we can also infer that the direction of these eruptions was toward the south-west for CMEs 1 and 4, and to the south for CMEs 2 and 5. As one can see from Figure~\ref{fig:source_region_eruptions}a--d, for all four events that originated from AR 13664, the flare ribbons (and the associated magnetic polarity inversion line (PIL) between them) have a low inclination (about 20$^\circ$ with respect to the solar equator). Considering several solar proxies, such as the flare ribbons (e.g. forward J-shape in Figure~\ref{fig:source_region_eruptions}a and d), the left-skew of overlying coronal loops (Figure~\ref{fig:handedness}a) and of post-eruptive flare arcades (Figure~\ref{fig:handedness}b), one would expect right-handed, low inclination, and hence south-west-north (SWN) flux ropes in situ, with the axial field direction pointing westward for all eruptive CMEs. Information about different solar proxies can be found in \citeA<e.g.>{palmerio2017determining}, and the definition of the flux rope types is given in \citeA{bothmer1998structure}.

CME 3 originated from a quiet-Sun filament eruption located to the south of AR 13667 and of the northern polar coronal hole, with its center around ($-25^\circ$E, 20$^\circ$N). The filament is visible in $\mathrm{H} \alpha$ observations of the GONG observatory in Cerro Tololo (Figure~\ref{fig:handedness}c). The extended erupting filament that caused CME 3 is associated with a distinct dimming (see Figure~\ref{fig:handedness}d), with its northern segment extending to the polar coronal hole. The location of this dimming and the right-skewed flare arcades connecting the dimming regions indicate a dextral filament which corresponds to a negative magnetic helicity \cite{martin1998,chen2014handedness}. The solar proxies would hence indicate a south-east-north (SEN) flux rope for CME 3 in situ.
          
\begin{table}[h]
    \caption{Flare parameters associated with the five CMEs that erupted on 8--9 May 2024. CME 3 is not linked to a flare, the peak flare time corresponds to the start of the filament eruption. CME 4 is associated with a double-peaked flare.}
    \centering
    \begin{tabular}{ccccc}
        \toprule
        CME & 1st obs. in LASCO/C2 & flare & source region & peak flare time \\      
        \midrule
        1 & 2024-05-08T05:36Z & X1.0 & S22W10 (13664) & 2024-05-08T05:09Z \\
        2 & 2024-05-08T12:24Z & M8.6 & S20W11 (13664) &  2024-05-08T12:04Z  \\
        3 & 2024-05-08T19:12Z & - & N26E22 (13667) & 2024-05-08T08:15Z \\
        \multirow{2}{*}{4} & \multirow{2}{*}{2024-05-08T22:24Z} & X1.0 & S20W17 (13664) & 2024-05-08T21:40Z  \\
         &  & M9.8 & S22W22 (13664) & 2024-05-08T22:27Z  \\
        5 & 2024-05-09T09:24Z & X2.2 & S20W25 (13664) & 2024-05-09T09:13Z  \\
        \bottomrule
    \end{tabular}
    \label{tab:initial_parameters_flare}
\end{table}

\begin{figure}[h!]
    \centering
    \includegraphics[width=\linewidth]{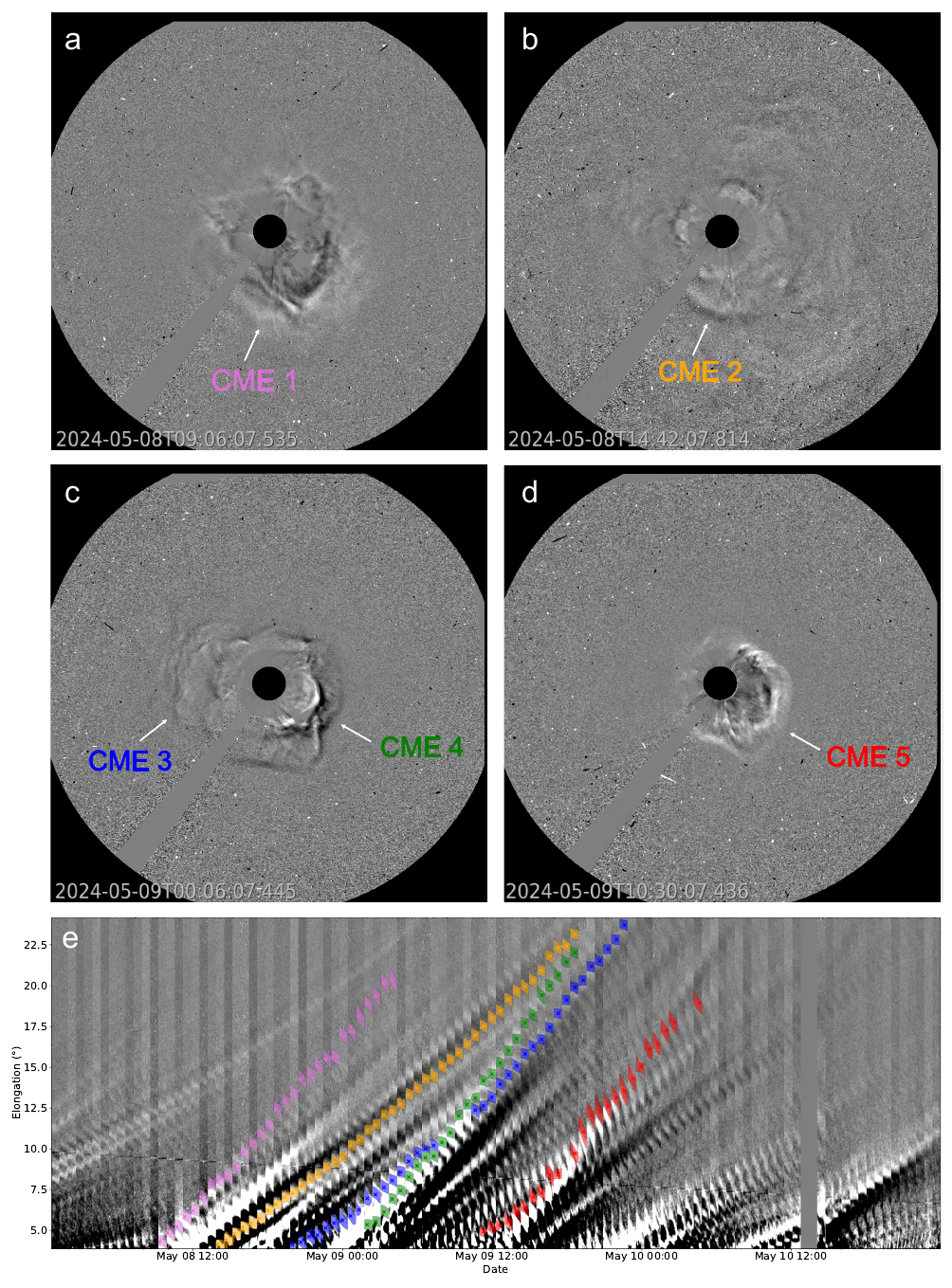}
    \caption{Interplanetary evolution of the five CMEs that caused the superstorm on 10--11 May 2024. (a)--(d) LASCO/C2 and C3 observations of the five halo CMEs with white arrows indicating the different fronts, (e) fronts of the different CMEs tracked in STEREO-A/HI1 running differences highlighted in the corresponding Jplot cut around the ecliptic.}
    \label{fig:interplanetary_evolution_hi}
\end{figure}

Panels a--d of Figure~\ref{fig:interplanetary_evolution_hi} show the five CMEs as observed with the Large Angle Spectroscopic Coronagraph \cite<LASCO;>{Brueckner_1995} on board the Solar and Heliospheric Observatory \cite<SOHO;>{domingo_1995}. The different CMEs are visible as halo CMEs and can be clearly identified separately. The first eruption enters the field of view (FOV) of LASCO/C2 on 8 May 2024, 05:36~UT (see also Table~\ref{tab:initial_parameters_flare}) and shows an increased brightness in a south-westerly direction. The second eruption is weaker than the first, but the front is still clearly distinguishable and appears for the first time at 12:24~UT in the LASCO/C2 FOV. Figure \ref{fig:interplanetary_evolution_hi}c shows CME 3 and CME 4: While the filament eruption from AR 13667 moves in a north-easterly direction, CME 4, which enters the field of view 3.37~hours later, propagates in a south-westerly direction. CME 5 is again visible as a halo CME starting on 9 May 2024, 09:24~UT (Figure~\ref{fig:interplanetary_evolution_hi}d). 

Figure~\ref{fig:interplanetary_evolution_hi}e displays the resulting Jplot from observations of the Heliospheric Imagers \cite<HI;>{eyles2009heliospheric} onboard STEREO-A that allow us to follow the propagation of the CMEs through the inner heliosphere. We identify the different fronts of the CMEs as indicated by the different colors, which also match the colored labels in Figures~\ref{fig:interplanetary_evolution_hi}a--d. We conclude that CME 3 is overtaken by CME 4 shortly after both enter the field of view of HI1. However, it is unclear whether the two CMEs physically interact. HI1 provides line-of-sight observations that could lead to an apparent overlap of these CMEs in the images, but this need not be the case in 3D. Considering that the longitudinal separation between the apices of CME 3 and CME 4 is quite large and amounts to approximately 32$^{\circ}$, we assume that CME 4 reaches and passes the heliocentric distances of CME 3 without any significant interaction occurring. The apex direction is taken from the Space Weather Database Of Notifications, Knowledge, Information (DONKI; \url{https://kauai.ccmc.gsfc.nasa.gov/DONKI/}) provided by the Moon to Mars (M2M) Space Weather Analysis Office and hosted by the Community Coordinated Modeling Center (CCMC; \url{https://ccmc.gsfc.nasa.gov/}). In this database, the kinematic properties of the CMEs are derived from coronagraph observations using the CME Analysis Tool of the Space Weather Prediction Center (SWPC CAT; {\url{https://ccmc.gsfc.nasa.gov/tools/SWPC-CAT/}}). Whenever various measurement types are specified in DONKI, e.g.\ tracking the shock front as well as the leading edge of a CME in the coronagraph images, we take the average of the resulting parameters. Movies containing observations of the five CMEs as observed in the SDO/AIA 171~{\AA}
filter, the LASCO/C2 and C3 coronagraphs, and the STEREO-A/HI1 heliospheric imager can be found in a figshare repository \cite<\url{https://figshare.com/articles/dataset/May2024Superstorm/27792873;}>{Weiler2024}.

\section{CME Arrival Time and Speed Modeling}\label{sec:elevo}

\begin{figure}[h!]
    \centering
    \includegraphics[width=\linewidth]{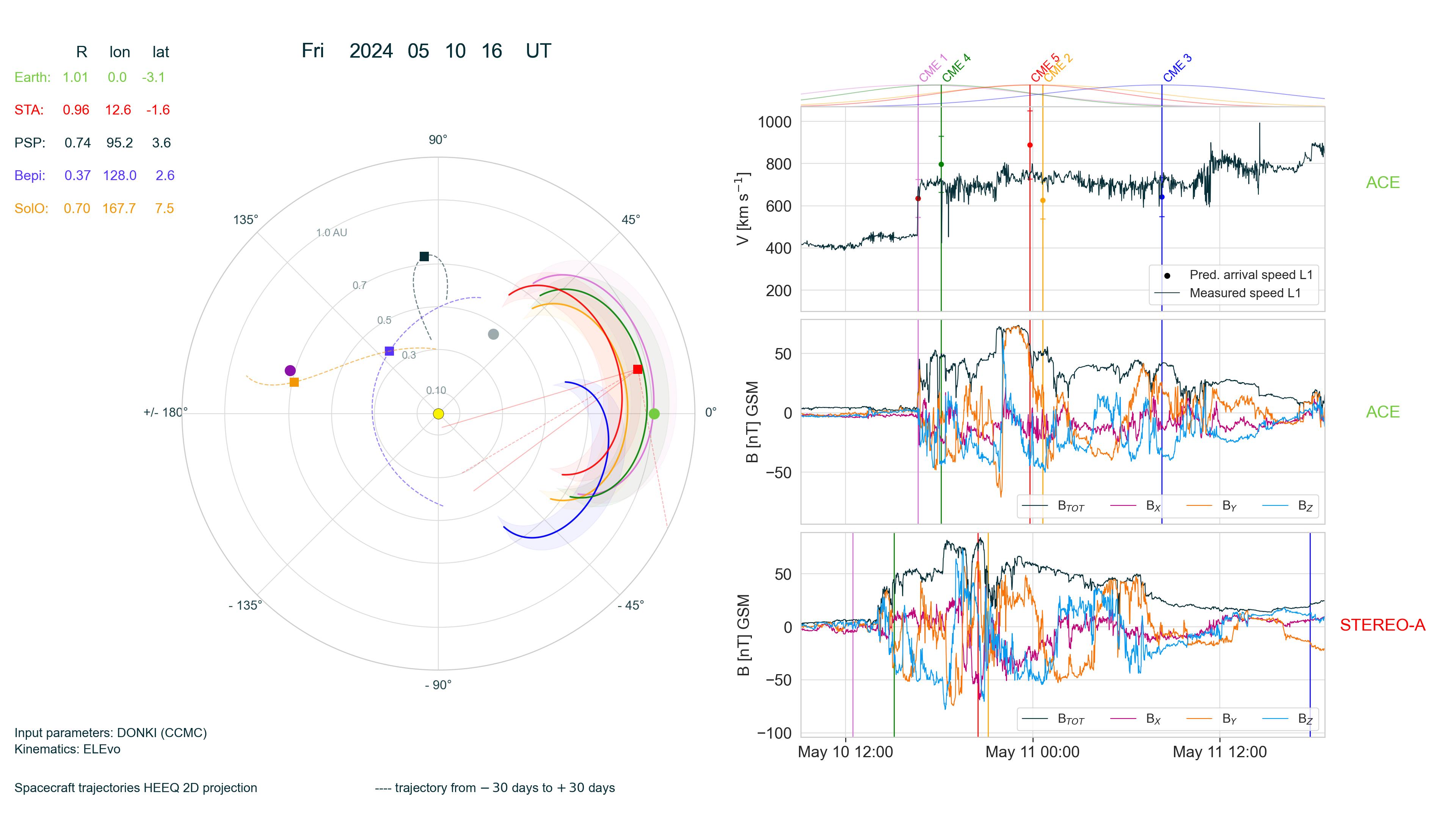}
    \caption{Snapshot of the ELEvo model at 10 May 2024, 16:00 UT. Left: A view of the heliosphere from above with the different spacecraft positions of STEREO-A (red), Solar Orbiter (SolO; orange), Parker Solar Probe (PSP; black), BepiColombo (Bepi; blue), and L1 (green) displayed in Heliocentric Earth Equatorial (HEEQ) coordinates. The propagation of the five different CMEs is shown, assuming an elliptical front, with the shaded areas indicating the $\pm 1 \sigma$ uncertainties of the arrival time. Right: In situ speed and magnetic field data from ACE in real-time, and STEREO-A real-time beacon data. The colored vertical lines indicate the predicted arrival times at L1 as issued by ELEvo, where the color-coding corresponds to the one in Figure~\ref{fig:interplanetary_evolution_hi}. The dots in the first panel correspond to the predicted arrival velocities at L1, with the error bars indicating the $\pm 1 \sigma$ uncertainty. Input parameters for the CMEs are taken from DONKI (CCMC).}
    \label{fig:interplanetary_evolution_elevo}
\end{figure}

Figure~\ref{fig:interplanetary_evolution_elevo} shows a snapshot of the ELliptical Evolution model \cite<ELEvo;>{moestl2015elevo} visualizing the propagation of the five CMEs through the heliosphere.

 The ELEvo model assumes an elliptical front for the CMEs and includes a simple drag-based model \cite{vrsnak2013propagation}. The snapshot is taken at the time of the measured arrival of the first CME at L1 on 8 May 2024 at 16:37~UT. On the left panel, a top-down view of the heliosphere with the different spacecraft and planet positions is presented in Heliocentric Earth Equatorial (HEEQ) coordinates. Real-time STEREO-A beacon and ACE data are shown on the right. The first panel shows the speed data from ACE. The second and third panels display the magnetic field data in Geocentric Solar Magnetospheric (GSM) coordinates from ACE and STEREO-A, respectively. Here we will discuss the simulation of the five CMEs with ELEvo, a detailed discussion of the in situ data follows in section~\ref{sec:in_situ}.

The CME parameters required as input for the ELEvo model, initial time at 21.5~R$_\odot$, initial speed and direction of the CME, are again taken from DONKI. As can be seen, ELEvo reproduces the actual arrival time of CME 1 at L1 well. A movie showing the full simulated propagation of the five CMEs is also available in \citeA{Weiler2024}. The time difference between the calculated and the observed arrival time is only $\sim$~2~min, with a model error window of $\pm 7$~hours. The model error results from including 100,000 ensemble members with slightly different values for the initial speed of the CME (taken from DONKI), the speed of the ambient solar wind and the drag parameter. The initial estimates for the ambient solar wind speed and the drag parameter are assumed to be 400~km\,s$^{-1}$ and $0.1\times10^{-7}$~km$^{-1}$, respectively. These parameters are then randomly varied at each time step along a normal distribution, with a standard deviation of 50~km\,s$^{-1}$ for both the initial speed of the CME and the speed of the ambient solar wind and $0.025\times10^{-7}$~km$^{-1}$ for the drag parameter \cite<see also>{calogovic2021dbem}. Atop the first in situ panel, we plot the normal distributions of the arrival time at L1 resulting from the ensemble for each CME. The shaded area around the ellipses on the left, simulating the shock fronts, shows the corresponding $\pm 1 \sigma$ uncertainty. In the speed panel, we also indicate the estimated arrival speed of the CMEs at L1, whereby the arrival speed is underestimated for CMEs 1, 2, and 3 and overestimated for CMEs 4 and 5. At 889~km\,s$^{-1}$, the arrival speed of CME 5, which is modeled to be the third to arrive at Earth, deviates the most from the actual measured speed of $\sim$700~km\,s$^{-1}$ at the predicted arrival time of CME 5. Although the ELEvo model predicts the arrival of the first CME at L1 accurately, it forecasts the arrival of the same CME at STEREO-A (see third panel) 1~h 34~min too early, which is, however, still a good prediction \cite<e.g.>[]{kay2024CMEarrivals}. 

The simulation shows that the apex direction of the CMEs originating from the southern hemisphere is slightly west of the Sun-Earth line. In particular, CME 1 with an inferred apex direction of 12$^\circ$ appears to be heading towards STEREO-A head-on. The filament eruption from the northern hemisphere (i.e.\ CME 3), in contrast, is moving more to the east and is the last to reach Earth, according to ELEvo. This is not necessarily the case for the actual propagation, as the model, which only propagates CME fronts, is not able to reproduce the interaction of the different CMEs, which could impact both their kinematics and directions. However, even by simply propagating the individual CMEs outwards, the simulation suggests that all five CMEs would arrive within 15.5~hours of each other at L1, which already hints at a rather complex in situ signature there. 

\section{In situ Spacecraft Observations}\label{sec:in_situ}

To study the in situ measurements of the five CMEs, we inspect real-time data from the ACE spacecraft, positioned at L1 (at 1.00~AU), and the STEREO-A spacecraft (at 0.96~AU) acting as a sub-L1 monitor. We use the ACE Magnetic Field Experiment \cite<MAG;>[]{smith1998ACEmag} and the Solar Wind Electron Proton Alpha Monitor \cite<SWEPAM;>[]{mccomas1998solar} for real-time measurements of the local (i.e. L1) magnetic field and solar wind plasma, respectively. The magnetometer instrument IMPACT \cite{acuna2008stereo} onboard STEREO-A is used for measurements of the magnetic field at the STEREO-A position (i.e. sub-L1). For our analysis, we use STEREO-A beacon magnetic field data, which is produced and available in real-time \cite{biesecker2008stereobeacon}. No real-time plasma data was available from STEREO-A at the time of the event.

\begin{figure}[h!]
    \centering
    \includegraphics[width=1.0\linewidth]{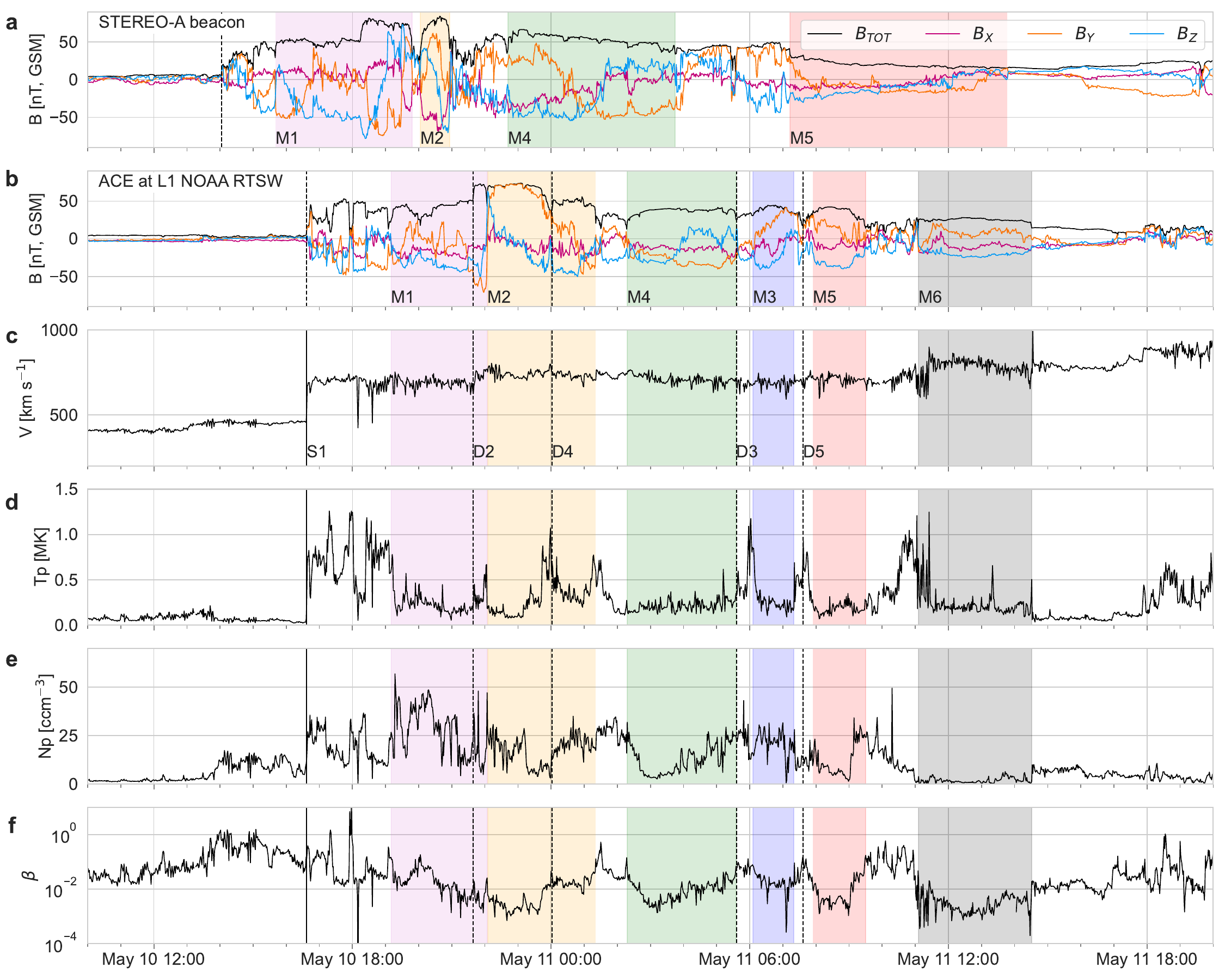}
    \caption{In situ measurements of the complex CME-CME interactions at L1 and STEREO-A. (a) The magnetic field vector observed in real-time by STEREO-A/IMPACT, located at 12.6$^\circ$ longitude west of Earth, with identified magnetic obstacles indicated as shaded regions and named M1, M2, M4, and M5. The first shock arrival time (10 May 2024 14:03 UT) is marked as a dashed line. (b), NOAA real-time solar wind magnetic field vector, provided by the magnetometer ACE/MAG. Identified magnetic obstacles are named M1--M6 and indicated by the corresponding shaded regions. The start times for each CME are given as dashed lines. (c) Plasma speed measured by ACE/SWEPAM. (d) Observations of the plasma proton temperature $T_p$ observed by the ACE/SWEPAM instrument. Enhancements of $T_p$ that we link to the individual CME arrivals are marked as vertical dashed lines named S1 for the initial shock, and then D2--D5 for discontinuities that mark the CME start times. (e) Solar wind plasma proton density $N_p$, and (f) plasma $\beta$, ratio of plasma to magnetic pressure, both from  ACE/SWEPAM.}
    \label{fig:interaction}
\end{figure}

Figure~\ref{fig:interaction}a shows the magnetic field data observed by STEREO-A IMPACT, where an interplanetary shock starts at 14:03 UT on 10 May 2024. The shock is preceded by a period of undisturbed solar wind conditions. Figures~\ref{fig:interaction}b--f show the magnetic field and plasma data measured by the ACE spacecraft at L1. In the magnetic field data of Figure~\ref{fig:interaction}b, an interplanetary shock starts on 10 May 2024 16:37 UT, 2~h 34~min after the shock was measured at STEREO-A. After the shock, six magnetic ejecta can be identified (M1--M6), which have interacted but not yet merged and can therefore still be individually distinguished. The identification of the different magnetic ejecta regions is mainly performed by focusing on rotations of the magnetic field vector and low plasma beta values (see Figure~\ref{fig:interaction}f). Due to the successive launch of the CMEs within 28~hours, the expansion of the individual CMEs is inhibited, leading to a preservation of high magnetic field strengths at L1, with a maximum of 74.8~nT on 10 May 2024, 23:05 UT. This is consistent with previous simulations of CME-CME interaction \cite<e.g.>{lugaz2005b,scolini2020}. Furthermore, the lack of expansion leads to a short duration of all ME, ranging from only 1h 14min for M3 to 3h 35min for M6. 

\begin{table}[h]
\caption{Positions of ACE and STEREO-A at shock arrival time in HEEQ coordinates, and arrival times of the shock and start and end times of the magnetic obstacles (UT) of the identified CMEs.}\label{tab:positions_times}
\begin{tabular*}{\textwidth}{@{\extracolsep\fill}ccccc}
\toprule%
        Spacecraft & shock arrival [UT] & R [au] & lon [deg] & lat [deg] \\
        \midrule
        ACE & 2024-05-10 16:37 & 1.000 & -0.013 & -3.072 \\
        STEREO-A & 2024-05-10 14:03 & 0.956 & 12.554  & -1.570 \\ 
        \midrule
         Spacecraft & CME & ICME start [UT] & MO start [UT] & MO end [UT] \\   
        \midrule
        ACE & 1 & 2024-05-10 16:37 & 2024-05-10 19:10  & 2024-05-10 22:04  \\
        ACE & 2 & 2024-05-10 21:38 & 2024-05-10 22:05 & 2024-05-11 01:20 \\
        ACE & 3 & 2024-05-11 05:36 & 2024-05-11 06:06 & 2024-05-11 07:20 \\
        ACE & 4 & 2024-05-11 00:02 & 2024-05-11 02:18 & 2024-05-11 05:35 \\
        ACE & 5 & 2024-05-11 07:37 & 2024-05-11 07:55 & 2024-05-11 09:30 \\  
        ACE & 6 & 2024-05-11 10:41 & 2024-05-11 11:06 & 2024-05-11 14:31 \\
        \midrule
        STEREO-A & 1 & 2024-05-10 14:03  & 2024-05-10 15:41 & 2024-05-10 19:48 \\  STEREO-A & 2 & 2024-05-10 18:14 & 2024-05-10 20:03 & 2024-05-10 20:56 \\
        STEREO-A & 4 & 2024-05-10 21:41 & 2024-05-10 22:42 & 2024-05-11 03:45 \\ 
        STEREO-A & 5 & 2024-05-11 05:36 & 2024-05-11 07:13  & 2024-05-11 13:46 \\
        \bottomrule
\end{tabular*}
\end{table}

\begin{figure}[h!]
    \centering
    \includegraphics[width=1.0\linewidth]{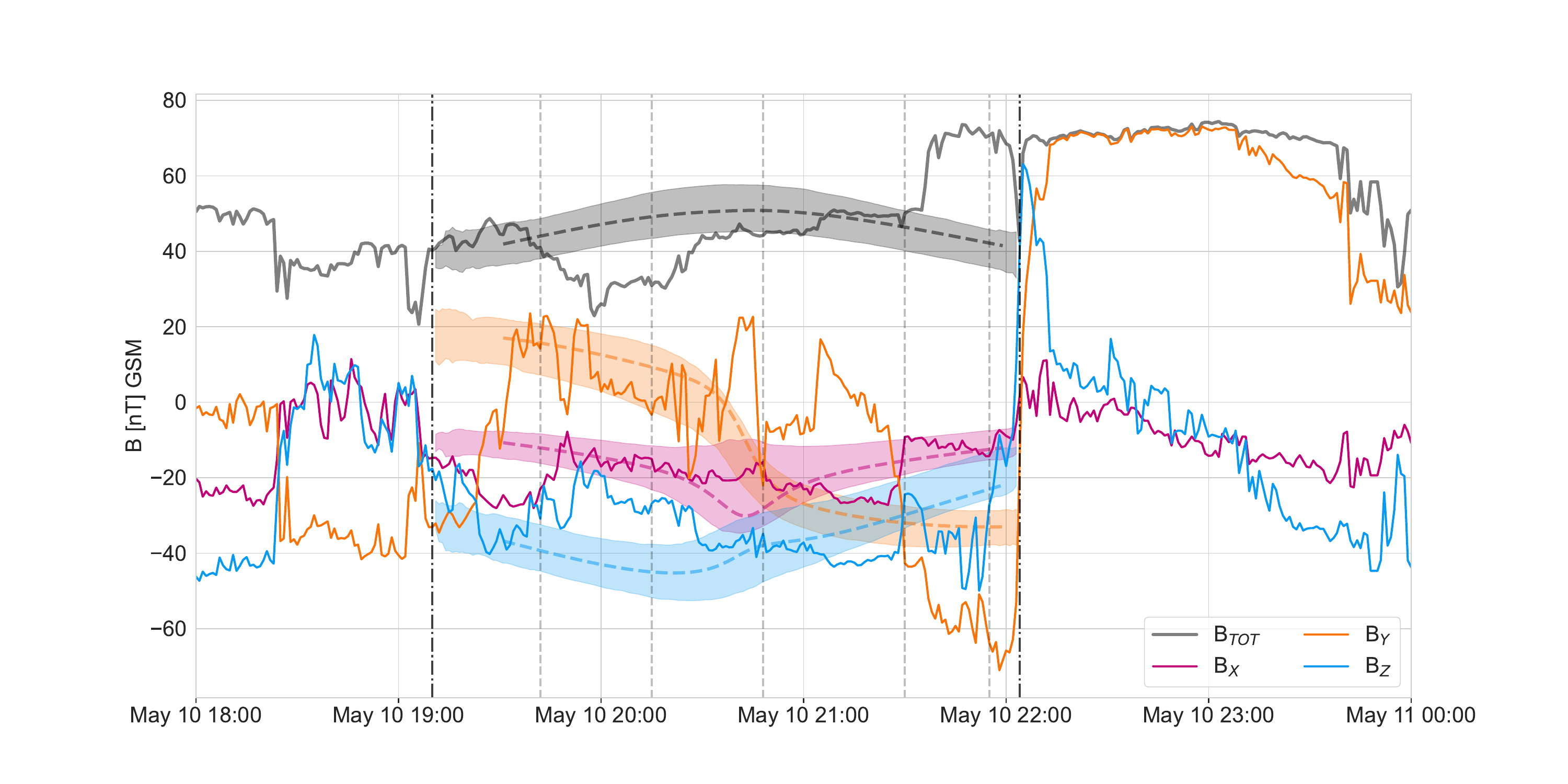}
    \caption{3DCORE reconstruction of the M1 in situ flux rope signature observed at L1. The shaded areas correspond to the $1 \sigma$ spread of the ensemble. The dashed colored lines show one specific flux rope from the ensemble. The vertical gray dashed lines delineate the fitting points at which 3DCORE evaluates the reconstruction, and the black dashed dotted lines show the start and end of the reconstruction interval. The data are presented in the GSM coordinate system.}
    \label{fig:3dcoreM1}
\end{figure}

Table~\ref{tab:positions_times} presents the corresponding times of the identified magnetic obstacles of the five magnetic ejecta (M1-M5), discussed in detail, as well as another magnetic flux rope (M6), which does not contribute as much to the geomagnetic response. For this reason, and also because no clear connection between the Sun and the in situ signature could be established in the absence of a corresponding LASCO observation, we do not include M6 in our analysis. Four out of the six ejecta identified in the ACE data are also identified in the STEREO-A data. More specifically, the similar rotational behavior of the flux ropes leads to the conclusion that M1, M2, M4 and M5 in the ACE data match the four identified ejecta in the STEREO-A data (as indicated by the colored shading). The ejecta identified in both STEREO-A and L1 data are right handed, with M2, M4, and M5 classified as north-east-south (NES), SWN, and east-south-west (ESW) flux rope types, respectively. The flux rope type of M1 is hard to determine from in situ data alone, however, when applying the 3D coronal rope ejection model (3DCORE) to the ACE data, the model indicates an ESW flux rope type (see Figure~\ref{fig:3dcoreM1}) as well as a right-handed, high inclination flux rope. The 3DCORE model assumes a Gold-Hoyle-like flux rope with an elliptical cross-section that stays attached to the Sun at all times. The tapered torus expands self-similarly and propagates through the heliosphere according to a drag-based model. Detailed information about 3DCORE can be found in \citeA{weiss2021analysis} and \citeA{weiss2021triple}.

Looking again at the source region, one would expect the CMEs emanating from AR 13664 to have a SWN flux rope in situ (see section~\ref{sec:solar_interplanetary_evolution}). This, however, is only valid for M4. M3 in the ACE data is the only flux rope that shows a left-handed helicity. More specifically, the flux rope type of M3 is SEN, which is consistent with the flux rope type inferred from solar proxies of AR 13667 in the northern hemisphere. This leads us to the conclusion that M3 corresponds to CME 3 associated with the filament eruption. In addition, M3 cannot be identified in the magnetic field data from STEREO-A, possibly due to the longitudinal separation of 34 degrees between AR 13667 and STEREO-A. In contrast, the in situ signatures M1, M2, M4, and M5 are most likely associated with CME 1, CME 2, CME 4, and CME 5, respectively, originating from AR 13664 in the southern hemisphere (see also Table~\ref{tab:positions_times}). Although the right- and left-handedness of each of the flux ropes is consistent with the solar proxies in the southern and northern hemispheres, respectively, we believe that the different flux ropes types in situ for M1, M2, and M5 are the result of the complex interplanetary evolution of the five successive CMEs, whose exact rotation and processes could not be fully deciphered. Furthermore, in \citeA{palmerio2018coronal}, the authors conducted a statistical analysis of the consistency of flux rope types between the Sun and Earth, finding a strict correspondence in only 20\% of the 20 events analyzed. Looking again at Figure~\ref{fig:interaction}a and b, it is noteworthy that the flux ropes originating from the southern hemisphere (i.~e. M1, M2, M4, M5) show periods of prolonged negative $B_z$, possibly resulting from the PIL orientation, which is tilted by 20$^{\circ}$ from the ecliptic plane, as discussed in section~\ref{sec:solar_interplanetary_evolution}, and in more detail in \citeA{wang2024}. 

We also search for plausible start times for the individual ICMEs, and identify discontinuities in the ACE data, where jumps in the magnetic field, density and temperature occur, and denote them as D2--D5 in Figure~\ref{fig:interaction}. We hypothesize that D2 was a shock driven by one of the overtaking CMEs and decayed into a compression wave due to the extreme magnetosonic speed $v_{ms} = 271$~km\,s$^{-1}$ inside the magnetic ejecta of M1. Nevertheless, D2 causes a compression of the upstream magnetic field magnitude, $B_{\mathrm{tot}}$, by a factor of 1.3, which also leads to an amplification of the $B_z$ component, with a minimum of -49~nT at 21:56~UT. D4 also has weak Alfvén and magnetosonic Mach numbers of $M_a=0.8$ and $M_{ms}=0.7$, respectively. It causes the $B_z$ component to drop from -2~nT to -36~nT, which remains increasingly negative for one hour. The parameters for these discontinuities as well as S1, which is a fast forward interplanetary shock, are given in Table~\ref{tab:shock_parameters}.We do not analyze D3 and D5 in more detail as they do not contribute as much to the geomagnetic effects. The shock/discontinuity normal is calculated using 10-min averaged up- and downstream values. For the discontinuity analysis, we use ACE science data instead of real-time data, as it provides the velocity components $v_x$, $v_y$, $v_z$. We can then use the velocity and magnetic field co-planarity method \cite{colburn1966} as well as a set of three equations for the mixed mode method \cite{abraham-shrauner1972,abraham-shrauner1976} to get five slightly different estimations of the shock/discontinuity normal. The shock speed is calculated from these estimates following \citeA{schwarz1998shock}. The resulting discontinuity normal varies the most for D4, which is why the uncertainty for the parameters derived thereof is large.

\begin{table}[h]
    \caption{Shock/discontinuity parameters for the features S1, D2 and D4. $c_s$, $v_a$, $v_{ms}$, correspond to sound, Alfvénic, and magnetosonic speed, respectively, $M_a$ is the Alfvénic and $M_{ms}$ the fast magnetosonic Mach number. The shock speed is given in the frame of the solar wind.}
    \centering
    \begin{tabular}{cccc}
        \toprule
         & S1 & D2 & D4 \\
        \midrule
        shock normal (deg) & 46 $\pm$ 6 & 81 $\pm$ 8 & 30 $\pm$ 29 \\
        shock speed (km\,s$^{-1}$) & 684 $\pm$ 41 & 854 $\pm$ 243 & 78 $\pm$ 16 \\ 
        upstream $v_a$ (km\,s$^{-1}$) & 267 & 240 & 226 \\ 
        upstream $c_s$ (km\,s$^{-1}$) & 46 & 48 & 74 \\
        upstream fast $v_{ms}$ (km\,s$^{-1}$) & 33 & 271 & 237 \\
        $M_a$ & 12.4 $\pm$ 0.4 & 1.1 $\pm$ 0.4 & 0.8 $\pm$ 0.7 \\
        $M_{ms}$ & 9.9 $\pm$ 0.3 & 1.1 $\pm$ 0.4 & 0.7 $\pm$ 0.6 \\
        \bottomrule
    \end{tabular}
    \label{tab:shock_parameters}
\end{table}

\section{Modeling of Geomagnetic Effects}\label{sec:results}


In this section, we demonstrate how a sub-L1 monitor may have been used to predict the geomagnetic effects of the May superstorm. To do this, we treat our analysis as though it was performed in real-time, meaning that we hindcast the geomagnetic effect using only data and knowledge that were available during the time of the event on 10 and 11 May. As previously stated, we aim to derive geomagnetic indices, namely $Dst$ and SYM-H, using STEREO-A beacon data, where SYM-H is the de facto high resolution $Dst$ index with a resolution of 1~minute \cite{wanliss2006}.

The geomagnetic effects are modeled with the TL model, which can be applied to any time resolution. When applied to data with hour and minute resolution, we compare the result with observed $Dst$ and SYM-H values, respectively, taken from low and high resolution OMNI data \cite<\url{https://spdf.gsfc.nasa.gov/pub/data/omni/};>[]{omnihour, omnimin}. 
The TL model requires the speed, density and magnetic field components in GSM coordinates of the solar wind at L1 as input. Since STEREO-A is 0.043~AU upstream of L1, the data has to be spatially and temporally mapped to L1 before the model can be applied. In addition, the TL model depends on the local time as well as past $Dst$ and magnetic field values. To initialize the model, estimates for the model parameters $Dst1$, $Dst2$, $Dst3$ are needed. 

\begin{figure}[h!]
    \centering
    \includegraphics[width=0.95\linewidth]{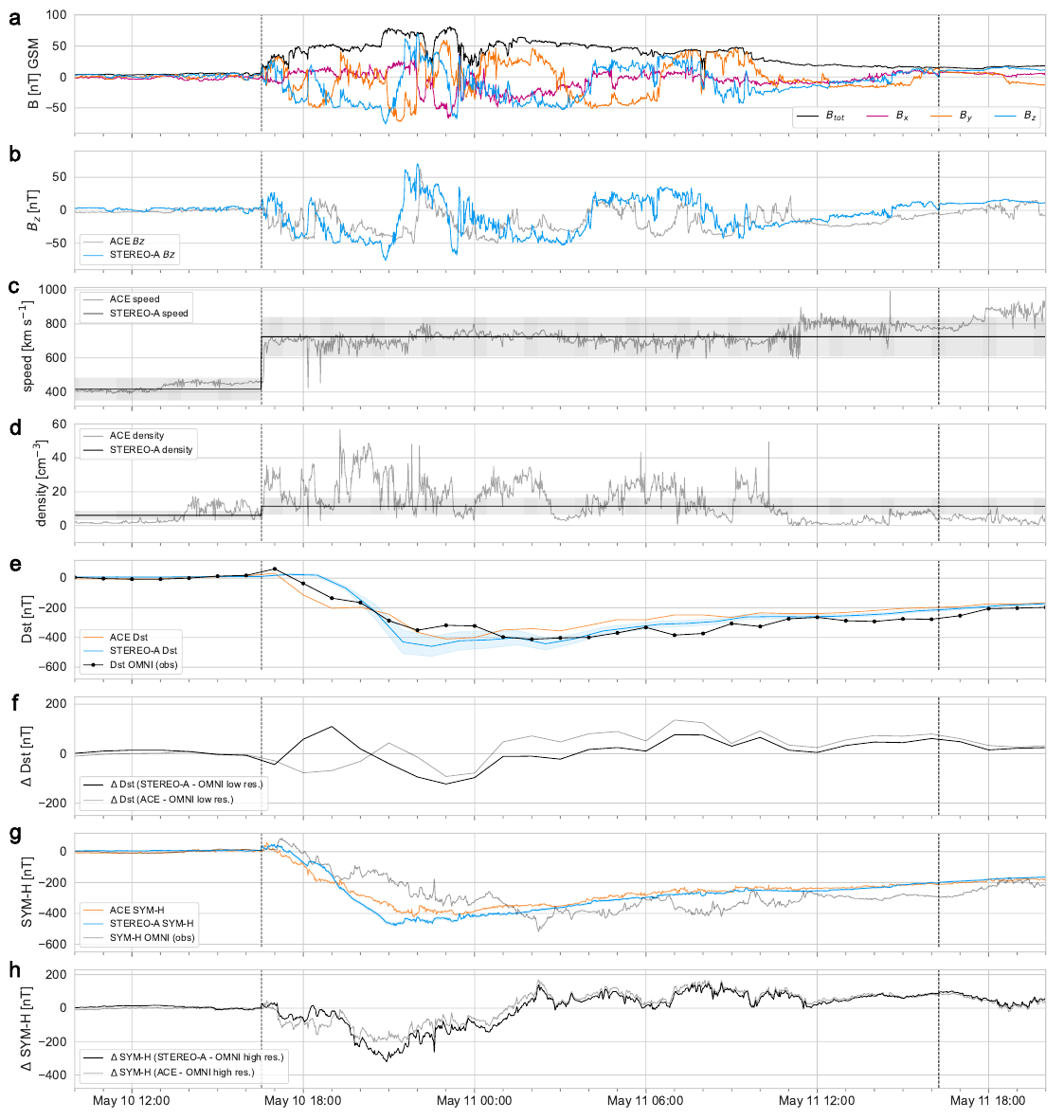}
    \caption{Modeling of the geomagnetic magnitude based on STEREO-A beacon data mapped to L1. (a) The scaled and shifted STEREO-A magnetic field components B$_{x}$,  B$_{y}$, B$_{z}$, and B$_{tot}$ in GSM coordinates, (b) shifted STEREO-A B$_{z}$ component compared to the B$_z$ component at L1, (c)-(d) estimated speed and density (black) compared to the bulk speed and density as measured by ACE at L1 (gray), (e) modeled $Dst$ index for STEREO-A (blue) and ACE (orange) data compared to the observed $Dst$ values as given in the low resolution OMNI dataset, (f) difference between observed and modeled $Dst$ index from STEREO-A (L1) data in black (gray), (g) Modeled SYM-H index for STEREO-A (blue) and ACE (orange) data compared to the observed SYM-H values as given in the high resolution OMNI dataset, (h) difference between observed and modeled SYM-H from STEREO-A (L1) data in black (gray). The vertical bars in each panel indicate the time range for which the error metrics are calculated.}
    \label{fig:dst_sta}
\end{figure}

Figure~\ref{fig:dst_sta} shows the mapped STEREO-A beacon data, compared to real-time ACE data, and the resulting geomagnetic indices calculated from the shifted STEREO-A as well as the observed L1 data. How the STEREO-A data is mapped to L1 and how missing values in the STEREO-A data are estimated is explained below.

We first transform the STEREO-A beacon data, which are available in radial-tangential-normal (RTN) coordinates, to the GSM coordinate frame by successively applying \citeA{hapgood1992coordinates} routines. The STEREO-A data are then mapped spatially and temporally to L1: Since STEREO-A is 0.043~AU closer to the Sun than L1, we expect the CMEs to expand further after being measured at STEREO-A and therefore to have a slightly lower magnetic field strength at L1. This assumption is based on previous statistical studies of global CME expansion, where CMEs are generally still expanding around 1~AU \cite<e.~g.>{davies2021catalogue}. Given the small radial separation of 0.043~AU between STEREO-A and L1, we expect a size expansion of only about 4\%. The expansion is accounted for by applying a power law $B_i(r) \propto r^{\alpha}, \,i = x, y, z$ with $\alpha=-1.66$ \cite{davies2021catalogue} to the STEREO-A magnetic field components $B_x$, $B_y$, $B_z$. Since ELEvo already indicates a highly interactive structure at L1, we consider the expansion rate an upper limit and also look at the case where there is no expansion at all (i.e. $\alpha=0$). These limits define the uncertainty for the magnetic field values, which is propagated when calculating the error for the geomagnetic indices. Figure~\ref{fig:dst_sta}a shows the scaled and temporally shifted STEREO-A magnetic field components.

As mentioned previously, STEREO-A beacon speed and density data were not available in real-time during the event. We therefore use the ELEvo model to estimate the different arrival speeds of the CMEs at STEREO-A. For the duration of the event (indicated by the vertical lines in Figure~\ref{fig:dst_sta}) we set the speed to $725~ \pm~ 116$~km\,s$^{-1}$, which is the average of all five predicted arrival speeds at STEREO-A. This agrees well with the observed mean speed at L1\,=\,716~km\,s$^{-1}$ during the event. The density during the event is estimated by using an average value derived from the HELIO4CAST ICMECAT catalog \cite{moestl2020ICMECAT}: The mean proton density for CMEs at L1 as measured by Wind is 11.5~cm$^{-3}$, including both the sheath region and magnetic obstacles. The speed and density before the event are estimated by averaging the ambient solar wind conditions at L1 in a 10-hour interval before the shock arrival at STEREO-A at 10 May 14:03 UT, which results in 416~km\,s$^{-1}$ and 6~cm$^{-3}$, respectively. Figures~\ref{fig:dst_sta}c--d show the estimated values for the speed and density at the STEREO-A position in comparison with measured speed and density values from ACE at L1.

We assume that the actual time difference between the shock measured at STEREO-A and L1 is unknown, again mimicking a real-time scenario. Instead, we estimate the arrival time using the averaged predicted arrival speed at STEREO-A, $725~\pm~116$~km\,s$^{-1}$, and the radial distance between STEREO-A and L1, 0.043~AU. The resulting time shift of $2.49~\pm~0.40$~hours is then applied to the whole STEREO-A dataset. With the actual time difference being 2.57~hours, we introduce an error of $\sim$~5~min to the geomagnetic index forecast. To differentiate between modeling and observational effects, we also calculate the geomagnetic indices using the observed L1 solar wind data and compare both results with the observed SYM-H and $Dst$ values.

To account for the error we make in shifting the data to L1, we randomly vary the scaled magnetic field components (within scaled and unscaled magnetic field values), our initial guesses for $Dst1$, $Dst2$, $Dst3$, the speed (within the arrival speed error), and density along a normal distribution using 100,000 ensemble members. We also randomly vary the time shift $\pm0.40$~hours, which corresponds to the error of the time shift. The random temporal shift also takes into account the different arrival times of the CMEs. Figures~\ref{fig:dst_sta}a--d show the mapped magnetic field data as well as our assumptions for speed and density with the shaded areas indicating $1 \sigma$ uncertainties. The resulting $Dst$ values for STEREO-A (blue) and L1 (orange) data are given in Figure~\ref{fig:dst_sta}e, which we can compare with the observed $Dst$ (black) and SYM-H (gray) values in Figure~\ref{fig:dst_sta}g. 

Table~\ref{tab:geomagnetic_indices} summarizes our results quantitatively and gives a comparison between observed and modeled geomagnetic indices as derived from shifted STEREO-A beacon and L1 data.

\begin{table}[h]
    \caption{Comparison between observed and modeled geomagnetic indices. Observed $Dst$ and SYM-H values are taken from low and high resolution OMNI datasets. The geomagnetic indices are modeled from mapped STEREO-A beacon data as well as L1 data using the TL model. The error represents the $1 \sigma$ uncertainty resulting from the ensemble.}
    \centering
    \begin{tabular}{cccccc}
        \toprule
        \multirow{2}{*}{Spacecraft} & Time $\mathrm{min}(Dst)$ & $\Delta t$ & $\mathrm{min}(Dst)$ & $\Delta \mathrm{min}(Dst)$ & RMSE($Dst$) \\
         & [UT] & [h] & [nT] & [nT] & [nT] \\
        \midrule
        OMNI & 2024-05-11 02:00 & -- & $-412.0$ & -- & -- \\
        STEREO-A & 2024-05-10 22:29 & $-3.5$ & $-459.9\pm54.8$ & $-47.9\pm55.4$ & $152.3\pm17.9$ \\
        L1 &  2024-05-10 23:00 & $-3.2$ & $-410.2$ & 1.8 & 71.2 \\
        \midrule
        \multirow{2}{*}{Spacecraft} & Time min(SYM-H) & $\Delta t$ & min(SYM-H) & $\Delta$min(SYM-H) & RMSE(SYM-H) \\
         & [UT] & [h] & [nT] & [nT] & [nT] \\
        \midrule
        OMNI. & 2024-05-11 02:14 & -- & $-518.0$ & -- & -- \\
        STEREO-A & 2024-05-10 21:13 & $-5.0$ & $-478.5\pm7.5$ & $39.5\pm9.7$ & $171.5\pm1.9$ \\
        L1 & 2024-05-10 21:53 & $-4.3$ & $-421.5$ & 96.5 & 94.9 \\
        \bottomrule
    \end{tabular}
    \label{tab:geomagnetic_indices}
\end{table}

As can be seen in Figure~\ref{fig:dst_sta}h, the largest differences between calculated and observed SYM-H values occur on 10 May at 21:13~UT, where the minimum SYM-H value from STEREO-A data is predicted 5.0~hours too early. It is noteworthy that the minimum $Dst$ and SYM-H values are also predicted too early by several hours when applying the TL model to L1 data. The minimum in $B_z$, associated with D2 as discussed in section~\ref{sec:in_situ}, occurs at approximately the same time as the predicted SYM-H minimum from L1 data on 10 May 21:53~UT. This $B_z$ minimum therefore appears to be an important driver for the TL model.  However, for the observed SYM-H, the prolonged period of $B_z$ after D4, which starts at 00:02~UT on 11 May, also appears to be a major contributor to the measured geomagnetic response. The actual SYM-H minimum is hence measured on 11 May 02:14~UT, whereas the model already predicts a recovery phase for indices derived from L1 and STEREO-A data. This leads to a rather high RMSE of 171.5~nT between STEREO-A and observed SYM-H values. The RMSE is calculated for the duration of the event only, where the duration is indicated by the vertical bars in Figure~\ref{fig:dst_sta}. Unsurprisingly, the observed $Dst$ and SYM-H values are better reproduced from L1 data, with RMSE = 71.2~nT and RMSE = 94.9~nT, respectively. However, due to the timing offset between calculated and modeled minimum $Dst$ and SYM-H, the error is still large.
When comparing the outcome from STEREO-A to SYM-H values derived using the L1 data, we calculate a RMSE of 39.2~nT, which is considerably lower than the RMSE of 171.5~nT calculated when comparing the STEREO-A SYM-H values to the observed OMNI SYM-H values. We therefore conclude that the model output is similar using either STEREO-A or L1 data, where the minima of the calculated geomagnetic indices are predicted earlier than the observed minima.

The observed minimum $Dst$ index = -412.0~nT is well reproduced by the L1 data, $Dst$ = -410.2~nT, while the measured minimum of SYM-H = -518.0~nT is underestimated by 23\%. Interestingly, the latter is also not resolved by the measured and hourly sampled $Dst$ index. The fact that the minimum $Dst$ value is better reproduced than the minimum SYM-H value is probably due to the bias we introduce when applying the TL model to 1-minute resolution data. The minimum $Dst$ value from STEREO-A data, on the other hand, overestimates the minimum intensity of the geomagnetic storm by 47.9~nT and underestimates the observed minimum SYM-H value by 39.5~nT, with the modeled minimum $Dst$ and SYM-H values being -459.9~nT and -478.5~nT, respectively.

\section{Conclusions}\label{sec:conclusion}

In this study, we have investigated the geomagnetic superstorm which occurred on 10--12 May 2024 and was measured by ACE and STEREO-A, among other spacecraft. The fortunate position of the STEREO-A spacecraft, 12.6 degrees away from the Sun-Earth line and 0.043~AU closer to the Sun resulted in an observation of the event two hours and 34 minutes earlier than at L1. The event consisted of five solar coronal mass ejections, launched from the Sun between 8--9 May 2024, that drove the main geomagnetic activity. This analysis gives unprecedented insights into how forecasting of highly interactive geomagnetic superstorms may work with future potential missions.

First, we give an overview of the event, where we analyze the source and in situ signatures at L1 separately and then explore their connection to better understand the cause for the high geomagnetic response. We are able to identify the source regions of the CMEs that cause the superstorm and the proxies thereof allow us to derive the handedness of the CMEs. With the CMEs from the northern (southern) hemisphere having negative (positive) helicity, they seem to follow the hemispheric polarity rule \cite{pevtsov2003helicityrule}. This also allows us to match the CMEs to their interplanetary counterparts as measured at L1 and STEREO-A. Furthermore, we model the propagation of the CMEs with the ELiptical Evolution model (ELEvo), which reproduces the arrival of the first CME at L1 well, with the difference between predicted and actual arrival being only 2~min, with a model error window of $\pm 7$~hours. In contrast, the arrival at STEREO-A is predicted 1~hour and 34~min too early. 
With the successive launch of these five CMEs within 28~hours, expansion of the individual CMEs seems to be inhibited, which results in a preservation of high magnetic fields strengths at L1. Interestingly, however, the CMEs have yet not fully merged at L1 and individual magnetic ejecta can still be recognized in the magnetic field data of STEREO-A and L1. It is noteworthy that the magnetic ejecta do not follow a global CME expansion law, with M1, M4, and M5 in particular being more extended at STEREO-A than at L1. We attribute this to the fact that the propagation directions of these CMEs, namely 12$^\circ$, 8.5$^\circ$, and 16.5$^\circ$ west of the Sun-Earth line, respectively, are directed more towards STEREO-A than L1. Consequently, in this case, the direction of propagation and longitudinal separation between STEREO-A and L1 seem to be the determining factors for the magnetic field strength rather than global expansion. Furthermore, we have identified discontinuities in the in situ data at L1 that appear to amplify $B_z$ and hence make an important contribution to the geomagnetic response. These $B_z$ minima could be the result of compressions possibly associated with CME-driven shocks propagating into the magnetic ejecta of the previous CMEs. Such shocks inside CMEs are frequent drivers of extreme $B_z$ and hence geomagnetic effects \cite<e.g.>{liu2014observations,lugaz2016shocks,scolini2020}.

For modeling the geomagnetic effect from STEREO-A data we only use data available at the time of the event, as we want to reproduce a real-time scenario. We therefore have to estimate the plasma parameters, density and speed, that are not available in real-time. The STEREO-A beacon data are then shifted temporally and spatially to L1 to determine the accuracy with which the strength of the geomagnetic superstorm could have been forecasted 2.57 hours earlier from a sub-L1 position. To this end, we apply the \citeA{Temerin2006} model to the mapped STEREO-A beacon data and compare the outcome to the observed $Dst$ and SYM-H values. While shifting the dataset to L1 requires only a few seconds, calculating the geomagnetic indices ($Dst$ and SYM-H) for STEREO-A, including 100,000 ensemble members, takes approximately 8 minutes. This is well below 4 -- 10~hours, which corresponds to the lead time for a CME propagating at 400~km\,s$^{-1}$ -- 1000~km\,s$^{-1}$ measured by a potential sub-L1 monitor at an orbital distance of 0.1 AU from Earth. Notably, a single run of the TL model takes less than a second, ensuring quick and efficient processing that is well-suited for real-time applications. The predicted $Dst$ and SYM-H minimum values of $-459.9\pm54.8$~nT and $-478.5\pm7.5$~nT, respectively, overestimate the observed minimum $Dst$ index of -412~nT by 11\% and underestimate the observed SYM-H of -518~nT by 8\%.\\ 
By also comparing the results calculated using STEREO-A data with the SYM-H values calculated from solar wind values measured at L1, we can isolate modeling effects from observational ones. We get an RMS error between the calculated SYM-H values from STEREO-A and L1 of 39.2~nT. When compared to the observed SYM-H values as given in the OMNI data, we get a comparatively high RMS error of 171.5~nT. The large RMSE most probably results from the minimum $Dst$ (SYM-H) value being forecasted 3.5 (5.0)~hours earlier than the observed minimum, which also leads to an early onset of the recovery phase. Since this is also the case for the $Dst$ and SYM-H values derived from L1 data, we argue that the temporal difference in the minima between model and observation is probably due to the TL model itself, while the difference in the actual strength of the geomagnetic impact is an observational effect, with the magnetic field strengths at STEREO-A being higher still.

In comparison, by combining the empirical methods from \citeA{burton1975} and \citeA{obrien2000}, the authors of \citeA{liu2024may} are able to reproduce the time of minimum $Dst$ index for this event better than the TL model. In this post-analysis, the authors also model the $Dst$ index of the May 2024 superstorm based on STEREO-A magnetic field data, but instead of estimating the missing plasma values, they use the density and solar wind speed as measured by the spacecraft \textit{Wind}. The minimum $Dst$ index calculated using L1 solar wind input in this paper is 8\% lower than the actual minimum and amounts to -378~nT. In contrast, the minimum $Dst$ index derived using STEREO-A data is -494~nT. The authors argue that the inferred geomagnetic indices from STEREO-A data give a lower limit, since the CMEs are more directed towards STEREO-A, possibly resulting in higher speeds and therefore even lower $Dst$ values there. We agree that the actual geomagnetic impact at STEREO-A should have been higher than our inferred $Dst$ of -459.9~nT, especially since our rough estimation of the density underestimates the measured mean density at L1 by 54\%. For a reliable derivation of the geomagnetic indices at a sub-L1 monitor position, it is therefore necessary that potential future missions are equipped with both magnetic field and plasma instruments. 

To summarize, we believe that this case study may represent a worst-case scenario for assessing the feasibility of future sub-L1 missions. The longitudinal distance between STEREO-A and L1 is at the outer range of what is recommended for sub-L1 missions \cite<below 15$^\circ$;>[]{good2016interplanetary,lugaz2024MEwidth,lugaz2024mission}, and the interaction of the five CMEs is both time- and space-dependent, with the time dependence being most evident in the dynamical evolution of D2. Despite these challenges, the prediction at STEREO-A is still acceptable compared to L1. Quantitatively, especially the period with SYM-H $< -200$~nT is predicted to be 21~h and 6~min, which agrees well with the measured duration of 23~h and 2~min. In addition, the overall correlation between modeled and observed geomagnetic indices is very good with a Pearson coefficient of 0.97. Probably not least because the RMSE between the $B_z$ component measured by STEREO-A and L1 is small and amounts to 26.7~nT, with $B_z$ being the most important parameter for the TL model. 
With our simple estimates, we would have been able to fairly accurately infer the strength of the geomagnetic storm 2.57~hours before its measurement at L1. This demonstrates the potential of future sub-L1 missions, especially since strongly interacting events, which are considered to be the main drivers of superstorms, cannot yet be reliably predicted otherwise. 

\section*{Data Availability Statement}



The data files used in this study and the animations that we produced can be found at \citeA{Weiler2024}.
The code for creating the paper figures and results is available at \citeA{weiler_2025_14772679}. The spacecraft data used in this study were obtained from these sources: STEREO-A beacon data were downloaded from the STEREO science center: \url{https://stereo-ssc.nascom.nasa.gov/data/beacon/ahead/impact/}. The ACE data are taken from the NOAA real time solar wind data product:  \url{https://services.swpc.noaa.gov/products/solar-wind/}. The OMNI dataset provides the geomagnetic disturbance storm time ($Dst$) index \cite{omnihour}, and the SYM-H index \cite{omnimin}. This research used version 6.0.2 of the SunPy open source software package \cite{sunpy_2020}.
This work made use of Astropy, a community-developed core Python package and an ecosystem of tools and resources for astronomy \cite{astropy:2022}.

\acknowledgments
We sincerely thank the reviewers for their valuable input and thoughtful feedback towards refining the structure of the paper and enhancing the clarity of key arguments. E.~W., C.~M., E.~D., U.~A., and H.~R. are funded by the European Union (ERC, HELIO4CAST, 101042188). Views and opinions expressed are however those of the author(s) only and do not necessarily reflect those of the European Union or the European Research Council Executive Agency. Neither the European Union nor the granting authority can be held responsible for them. We acknowledge the Community Coordinated Modeling Center (CCMC) at Goddard Space Flight Center for the use of the Space Weather Database Of Notifications, Knowledge, Information (DONKI), \url{https://kauai.ccmc.gsfc.nasa.gov/DONKI/}.
N.~L. acknowledges funding from 80NSSC24K1245 and 80NSSC20K0431. T.~A., J.~L. and M.~B. acknowledge that this research was funded in whole, or in part, by the Austrian Science Fund (FWF) [P 36093]. S.~M. and M.~R. acknowledge that this research was funded in whole, or in part, by the Austrian Science Fund (FWF) [P 34437]. A.M.V. acknowledges the Austrian Science Fund (FWF), project no.\ 10.55776/PAT7894023. For the purpose of open access, the author has applied a CC BY public copyright licence to any Author Accepted Manuscript version arising from this submission.

%
\bibliography{bibliography} 

\begin{thebibliography}{}

\bibitem [\protect \citeauthoryear {%
Abraham-Shrauner%
}{%
Abraham-Shrauner%
}{%
{\protect \APACyear {1972}}%
}]{%
abraham-shrauner1972}
\APACinsertmetastar {%
abraham-shrauner1972}%
\begin{APACrefauthors}%
Abraham-Shrauner, B.%
\end{APACrefauthors}%
\unskip\
\newblock
\APACrefYearMonthDay{1972}{}{}.
\newblock
{\BBOQ}\APACrefatitle {Determination of magnetohydrodynamic shock normals} {Determination of magnetohydrodynamic shock normals}.{\BBCQ}
\newblock
\APACjournalVolNumPages{Journal of Geophysical Research (1896-1977)}{77}{4}{736-739}.
\newblock
\begin{APACrefURL} \url{https://agupubs.onlinelibrary.wiley.com/doi/abs/10.1029/JA077i004p00736} \end{APACrefURL}
\newblock
\begin{APACrefDOI} \doi{https://doi.org/10.1029/JA077i004p00736} \end{APACrefDOI}
\PrintBackRefs{\CurrentBib}

\bibitem [\protect \citeauthoryear {%
Abraham-Shrauner%
\ \BBA {} Yun%
}{%
Abraham-Shrauner%
\ \BBA {} Yun%
}{%
{\protect \APACyear {1976}}%
}]{%
abraham-shrauner1976}
\APACinsertmetastar {%
abraham-shrauner1976}%
\begin{APACrefauthors}%
Abraham-Shrauner, B.%
\BCBT {}\ \BBA {} Yun, S\BPBI H.%
\end{APACrefauthors}%
\unskip\
\newblock
\APACrefYearMonthDay{1976}{}{}.
\newblock
{\BBOQ}\APACrefatitle {Interplanetary shocks seen by Ames Plasma Probe on Pioneer 6 and 7} {Interplanetary shocks seen by ames plasma probe on pioneer 6 and 7}.{\BBCQ}
\newblock
\APACjournalVolNumPages{Journal of Geophysical Research (1896-1977)}{81}{13}{2097-2102}.
\newblock
\begin{APACrefURL} \url{https://agupubs.onlinelibrary.wiley.com/doi/abs/10.1029/JA081i013p02097} \end{APACrefURL}
\newblock
\begin{APACrefDOI} \doi{https://doi.org/10.1029/JA081i013p02097} \end{APACrefDOI}
\PrintBackRefs{\CurrentBib}

\bibitem [\protect \citeauthoryear {%
{Acu{\~n}a}%
\ \protect \BOthers {.}}{%
{Acu{\~n}a}%
\ \protect \BOthers {.}}{%
{\protect \APACyear {2008}}%
}]{%
acuna2008stereo}
\APACinsertmetastar {%
acuna2008stereo}%
\begin{APACrefauthors}%
{Acu{\~n}a}, M\BPBI H.%
, {Curtis}, D.%
, {Scheifele}, J\BPBI L.%
, {Russell}, C\BPBI T.%
, {Schroeder}, P.%
, {Szabo}, A.%
\BCBL {}\ \BBA {} {Luhmann}, J\BPBI G.%
\end{APACrefauthors}%
\unskip\
\newblock
\APACrefYearMonthDay{2008}{{\APACmonth{04}}}{}.
\newblock
{\BBOQ}\APACrefatitle {{The STEREO/IMPACT Magnetic Field Experiment}} {{The STEREO/IMPACT Magnetic Field Experiment}}.{\BBCQ}
\newblock
\APACjournalVolNumPages{\ssr}{136}{1-4}{203-226}.
\newblock
\begin{APACrefDOI} \doi{10.1007/s11214-007-9259-2} \end{APACrefDOI}
\PrintBackRefs{\CurrentBib}

\bibitem [\protect \citeauthoryear {%
{Astropy Collaboration}%
\ \protect \BOthers {.}}{%
{Astropy Collaboration}%
\ \protect \BOthers {.}}{%
{\protect \APACyear {2022}}%
}]{%
astropy:2022}
\APACinsertmetastar {%
astropy:2022}%
\begin{APACrefauthors}%
{Astropy Collaboration}%
, {Price-Whelan}, A\BPBI M.%
, {Lim}, P\BPBI L.%
, {Earl}, N.%
, {Starkman}, N.%
, {Bradley}, L.%
\BDBL {}{Astropy Project Contributors}%
\end{APACrefauthors}%
\unskip\
\newblock
\APACrefYearMonthDay{2022}{{\APACmonth{08}}}{}.
\newblock
{\BBOQ}\APACrefatitle {{The Astropy Project: Sustaining and Growing a Community-oriented Open-source Project and the Latest Major Release (v5.0) of the Core Package}} {{The Astropy Project: Sustaining and Growing a Community-oriented Open-source Project and the Latest Major Release (v5.0) of the Core Package}}.{\BBCQ}
\newblock
\APACjournalVolNumPages{\apj}{935}{2}{167}.
\newblock
\begin{APACrefURL} \url{http://www.astropy.org} \end{APACrefURL}
\newblock
\begin{APACrefDOI} \doi{10.3847/1538-4357/ac7c74} \end{APACrefDOI}
\PrintBackRefs{\CurrentBib}

\bibitem [\protect \citeauthoryear {%
{Biesecker}%
, {Webb}%
\BCBL {}\ \BBA {} {St. Cyr}%
}{%
{Biesecker}%
\ \protect \BOthers {.}}{%
{\protect \APACyear {2008}}%
}]{%
biesecker2008stereobeacon}
\APACinsertmetastar {%
biesecker2008stereobeacon}%
\begin{APACrefauthors}%
{Biesecker}, D\BPBI A.%
, {Webb}, D\BPBI F.%
\BCBL {}\ \BBA {} {St. Cyr}, O\BPBI C.%
\end{APACrefauthors}%
\unskip\
\newblock
\APACrefYearMonthDay{2008}{{\APACmonth{04}}}{}.
\newblock
{\BBOQ}\APACrefatitle {{STEREO Space Weather and the Space Weather Beacon}} {{STEREO Space Weather and the Space Weather Beacon}}.{\BBCQ}
\newblock
\APACjournalVolNumPages{\ssr}{136}{1-4}{45-65}.
\newblock
\begin{APACrefDOI} \doi{10.1007/s11214-007-9165-7} \end{APACrefDOI}
\PrintBackRefs{\CurrentBib}

\bibitem [\protect \citeauthoryear {%
Borovsky%
}{%
Borovsky%
}{%
{\protect \APACyear {2018}}%
}]{%
Borovsky2018}
\APACinsertmetastar {%
Borovsky2018}%
\begin{APACrefauthors}%
Borovsky, J\BPBI E.%
\end{APACrefauthors}%
\unskip\
\newblock
\APACrefYearMonthDay{2018}{}{}.
\newblock
{\BBOQ}\APACrefatitle {The spatial structure of the oncoming solar wind at Earth and the shortcomings of a solar-wind monitor at L1} {The spatial structure of the oncoming solar wind at earth and the shortcomings of a solar-wind monitor at l1}.{\BBCQ}
\newblock
\APACjournalVolNumPages{Journal of Atmospheric and Solar-Terrestrial Physics}{177}{}{2-11}.
\newblock
\begin{APACrefURL} \url{https://www.sciencedirect.com/science/article/pii/S1364682617300159} \end{APACrefURL}
\newblock
\APACrefnote{Dynamics of the Sun-Earth System: Recent Observations and Predictions}
\newblock
\begin{APACrefDOI} \doi{https://doi.org/10.1016/j.jastp.2017.03.014} \end{APACrefDOI}
\PrintBackRefs{\CurrentBib}

\bibitem [\protect \citeauthoryear {%
{Bothmer}%
\ \BBA {} {Schwenn}%
}{%
{Bothmer}%
\ \BBA {} {Schwenn}%
}{%
{\protect \APACyear {1998}}%
}]{%
bothmer1998structure}
\APACinsertmetastar {%
bothmer1998structure}%
\begin{APACrefauthors}%
{Bothmer}, V.%
\BCBT {}\ \BBA {} {Schwenn}, R.%
\end{APACrefauthors}%
\unskip\
\newblock
\APACrefYearMonthDay{1998}{{\APACmonth{01}}}{}.
\newblock
{\BBOQ}\APACrefatitle {{The structure and origin of magnetic clouds in the solar wind}} {{The structure and origin of magnetic clouds in the solar wind}}.{\BBCQ}
\newblock
\APACjournalVolNumPages{\angeo}{16}{}{1-24}.
\newblock
\begin{APACrefDOI} \doi{10.1007/s00585-997-0001-x} \end{APACrefDOI}
\PrintBackRefs{\CurrentBib}

\bibitem [\protect \citeauthoryear {%
Boynton%
, Balikhin%
, Billings%
, Sharma%
\BCBL {}\ \BBA {} Amariutei%
}{%
Boynton%
\ \protect \BOthers {.}}{%
{\protect \APACyear {2011}}%
}]{%
boynton2011}
\APACinsertmetastar {%
boynton2011}%
\begin{APACrefauthors}%
Boynton, R\BPBI J.%
, Balikhin, M\BPBI A.%
, Billings, S\BPBI A.%
, Sharma, A\BPBI S.%
\BCBL {}\ \BBA {} Amariutei, O\BPBI A.%
\end{APACrefauthors}%
\unskip\
\newblock
\APACrefYearMonthDay{2011}{}{}.
\newblock
{\BBOQ}\APACrefatitle {Data derived NARMAX Dst model} {Data derived narmax dst model}.{\BBCQ}
\newblock
\APACjournalVolNumPages{Annales Geophysicae}{29}{6}{965--971}.
\newblock
\begin{APACrefURL} \url{https://angeo.copernicus.org/articles/29/965/2011/} \end{APACrefURL}
\newblock
\begin{APACrefDOI} \doi{10.5194/angeo-29-965-2011} \end{APACrefDOI}
\PrintBackRefs{\CurrentBib}

\bibitem [\protect \citeauthoryear {%
{Brueckner}%
\ \protect \BOthers {.}}{%
{Brueckner}%
\ \protect \BOthers {.}}{%
{\protect \APACyear {1995}}%
}]{%
Brueckner_1995}
\APACinsertmetastar {%
Brueckner_1995}%
\begin{APACrefauthors}%
{Brueckner}, G\BPBI E.%
, {Howard}, R\BPBI A.%
, {Koomen}, M\BPBI J.%
, {Korendyke}, C\BPBI M.%
, {Michels}, D\BPBI J.%
, {Moses}, J\BPBI D.%
\BDBL {}{Eyles}, C\BPBI J.%
\end{APACrefauthors}%
\unskip\
\newblock
\APACrefYearMonthDay{1995}{{\APACmonth{12}}}{}.
\newblock
{\BBOQ}\APACrefatitle {{The Large Angle Spectroscopic Coronagraph (LASCO)}} {{The Large Angle Spectroscopic Coronagraph (LASCO)}}.{\BBCQ}
\newblock
\APACjournalVolNumPages{\solphys}{162}{}{357-402}.
\newblock
\begin{APACrefDOI} \doi{10.1007/BF00733434} \end{APACrefDOI}
\PrintBackRefs{\CurrentBib}

\bibitem [\protect \citeauthoryear {%
Burt%
\ \BBA {} Smith%
}{%
Burt%
\ \BBA {} Smith%
}{%
{\protect \APACyear {2012}}%
}]{%
dscovr}
\APACinsertmetastar {%
dscovr}%
\begin{APACrefauthors}%
Burt, J.%
\BCBT {}\ \BBA {} Smith, B.%
\end{APACrefauthors}%
\unskip\
\newblock
\APACrefYearMonthDay{2012}{}{}.
\newblock
{\BBOQ}\APACrefatitle {Deep Space Climate Observatory: The DSCOVR mission} {Deep space climate observatory: The dscovr mission}.{\BBCQ}
\newblock
\BIn{} \APACrefbtitle {2012 IEEE Aerospace Conference} {2012 ieee aerospace conference}\ (\BPG~1-13).
\newblock
\begin{APACrefDOI} \doi{10.1109/AERO.2012.6187025} \end{APACrefDOI}
\PrintBackRefs{\CurrentBib}

\bibitem [\protect \citeauthoryear {%
Burton%
, McPherron%
\BCBL {}\ \BBA {} Russell%
}{%
Burton%
\ \protect \BOthers {.}}{%
{\protect \APACyear {1975}}%
}]{%
burton1975}
\APACinsertmetastar {%
burton1975}%
\begin{APACrefauthors}%
Burton, R\BPBI K.%
, McPherron, R\BPBI L.%
\BCBL {}\ \BBA {} Russell, C\BPBI T.%
\end{APACrefauthors}%
\unskip\
\newblock
\APACrefYearMonthDay{1975}{}{}.
\newblock
{\BBOQ}\APACrefatitle {An empirical relationship between interplanetary conditions and Dst} {An empirical relationship between interplanetary conditions and dst}.{\BBCQ}
\newblock
\APACjournalVolNumPages{Journal of Geophysical Research (1896-1977)}{80}{31}{4204-4214}.
\newblock
\begin{APACrefURL} \url{https://agupubs.onlinelibrary.wiley.com/doi/abs/10.1029/JA080i031p04204} \end{APACrefURL}
\newblock
\begin{APACrefDOI} \doi{https://doi.org/10.1029/JA080i031p04204} \end{APACrefDOI}
\PrintBackRefs{\CurrentBib}

\bibitem [\protect \citeauthoryear {%
{Chen}%
, {Harra}%
\BCBL {}\ \BBA {} {Fang}%
}{%
{Chen}%
\ \protect \BOthers {.}}{%
{\protect \APACyear {2014}}%
}]{%
chen2014handedness}
\APACinsertmetastar {%
chen2014handedness}%
\begin{APACrefauthors}%
{Chen}, P\BPBI F.%
, {Harra}, L\BPBI K.%
\BCBL {}\ \BBA {} {Fang}, C.%
\end{APACrefauthors}%
\unskip\
\newblock
\APACrefYearMonthDay{2014}{{\APACmonth{03}}}{}.
\newblock
{\BBOQ}\APACrefatitle {{Imaging and Spectroscopic Observations of a Filament Channel and the Implications for the Nature of Counter-streamings}} {{Imaging and Spectroscopic Observations of a Filament Channel and the Implications for the Nature of Counter-streamings}}.{\BBCQ}
\newblock
\APACjournalVolNumPages{\apj}{784}{1}{50}.
\newblock
\begin{APACrefDOI} \doi{10.1088/0004-637X/784/1/50} \end{APACrefDOI}
\PrintBackRefs{\CurrentBib}

\bibitem [\protect \citeauthoryear {%
{Colburn}%
\ \BBA {} {Sonett}%
}{%
{Colburn}%
\ \BBA {} {Sonett}%
}{%
{\protect \APACyear {1966}}%
}]{%
colburn1966}
\APACinsertmetastar {%
colburn1966}%
\begin{APACrefauthors}%
{Colburn}, D\BPBI S.%
\BCBT {}\ \BBA {} {Sonett}, C\BPBI P.%
\end{APACrefauthors}%
\unskip\
\newblock
\APACrefYearMonthDay{1966}{{\APACmonth{06}}}{}.
\newblock
{\BBOQ}\APACrefatitle {{Discontinuities in the Solar Wind}} {{Discontinuities in the Solar Wind}}.{\BBCQ}
\newblock
\APACjournalVolNumPages{\ssr}{5}{4}{439-506}.
\newblock
\begin{APACrefDOI} \doi{10.1007/BF00240575} \end{APACrefDOI}
\PrintBackRefs{\CurrentBib}

\bibitem [\protect \citeauthoryear {%
Collado~Villaverde%
, Muñoz%
\BCBL {}\ \BBA {} Cid%
}{%
Collado~Villaverde%
\ \protect \BOthers {.}}{%
{\protect \APACyear {2023}}%
}]{%
villaverde2023}
\APACinsertmetastar {%
villaverde2023}%
\begin{APACrefauthors}%
Collado~Villaverde, A.%
, Muñoz, P.%
\BCBL {}\ \BBA {} Cid, C.%
\end{APACrefauthors}%
\unskip\
\newblock
\APACrefYearMonthDay{2023}{10}{}.
\newblock
{\BBOQ}\APACrefatitle {Classifying and bounding geomagnetic storms based on the SYM-H and ASY-H indices} {Classifying and bounding geomagnetic storms based on the sym-h and asy-h indices}.{\BBCQ}
\newblock
\APACjournalVolNumPages{Natural Hazards}{120}{}{1-22}.
\newblock
\begin{APACrefDOI} \doi{10.1007/s11069-023-06241-1} \end{APACrefDOI}
\PrintBackRefs{\CurrentBib}

\bibitem [\protect \citeauthoryear {%
{Cyr}%
\ \protect \BOthers {.}}{%
{Cyr}%
\ \protect \BOthers {.}}{%
{\protect \APACyear {2000}}%
}]{%
stcyr2000diamond}
\APACinsertmetastar {%
stcyr2000diamond}%
\begin{APACrefauthors}%
{Cyr}, O\BPBI C\BPBI S.%
, {Mesarch}, M\BPBI A.%
, {Maldonado}, H\BPBI M.%
, {Folta}, D\BPBI C.%
, {Harper}, A\BPBI D.%
, {Davila}, J\BPBI M.%
\BCBL {}\ \BBA {} {Fisher}, R\BPBI R.%
\end{APACrefauthors}%
\unskip\
\newblock
\APACrefYearMonthDay{2000}{{\APACmonth{09}}}{}.
\newblock
{\BBOQ}\APACrefatitle {{Space Weather Diamond: a four spacecraft monitoring system}} {{Space Weather Diamond: a four spacecraft monitoring system}}.{\BBCQ}
\newblock
\APACjournalVolNumPages{Journal of Atmospheric and Solar-Terrestrial Physics}{62}{14}{1251-1255}.
\newblock
\begin{APACrefDOI} \doi{10.1016/S1364-6826(00)00069-9} \end{APACrefDOI}
\PrintBackRefs{\CurrentBib}

\bibitem [\protect \citeauthoryear {%
{Davies}%
, {Forsyth}%
, {Winslow}%
, {M{\"o}stl}%
\BCBL {}\ \BBA {} {Lugaz}%
}{%
{Davies}%
\ \protect \BOthers {.}}{%
{\protect \APACyear {2021}}%
}]{%
davies2021catalogue}
\APACinsertmetastar {%
davies2021catalogue}%
\begin{APACrefauthors}%
{Davies}, E\BPBI E.%
, {Forsyth}, R\BPBI J.%
, {Winslow}, R\BPBI M.%
, {M{\"o}stl}, C.%
\BCBL {}\ \BBA {} {Lugaz}, N.%
\end{APACrefauthors}%
\unskip\
\newblock
\APACrefYearMonthDay{2021}{{\APACmonth{12}}}{}.
\newblock
{\BBOQ}\APACrefatitle {{A Catalog of Interplanetary Coronal Mass Ejections Observed by Juno between 1 and 5.4 au}} {{A Catalog of Interplanetary Coronal Mass Ejections Observed by Juno between 1 and 5.4 au}}.{\BBCQ}
\newblock
\APACjournalVolNumPages{\apj}{923}{2}{136}.
\newblock
\begin{APACrefDOI} \doi{10.3847/1538-4357/ac2ccb} \end{APACrefDOI}
\PrintBackRefs{\CurrentBib}

\bibitem [\protect \citeauthoryear {%
{Domingo}%
, {Fleck}%
\BCBL {}\ \BBA {} {Poland}%
}{%
{Domingo}%
\ \protect \BOthers {.}}{%
{\protect \APACyear {1995}}%
}]{%
domingo_1995}
\APACinsertmetastar {%
domingo_1995}%
\begin{APACrefauthors}%
{Domingo}, V.%
, {Fleck}, B.%
\BCBL {}\ \BBA {} {Poland}, A\BPBI I.%
\end{APACrefauthors}%
\unskip\
\newblock
\APACrefYearMonthDay{1995}{{\APACmonth{12}}}{}.
\newblock
{\BBOQ}\APACrefatitle {{The SOHO Mission: an Overview}} {{The SOHO Mission: an Overview}}.{\BBCQ}
\newblock
\APACjournalVolNumPages{\solphys}{162}{1-2}{1-37}.
\newblock
\begin{APACrefDOI} \doi{10.1007/BF00733425} \end{APACrefDOI}
\PrintBackRefs{\CurrentBib}

\bibitem [\protect \citeauthoryear {%
{Eastwood}%
, {Kataria}%
, {McInnes}%
, {Barnes}%
\BCBL {}\ \BBA {} {Mulligan}%
}{%
{Eastwood}%
\ \protect \BOthers {.}}{%
{\protect \APACyear {2015}}%
}]{%
eastwood2015}
\APACinsertmetastar {%
eastwood2015}%
\begin{APACrefauthors}%
{Eastwood}, J\BPBI P.%
, {Kataria}, D\BPBI O.%
, {McInnes}, C\BPBI R.%
, {Barnes}, N\BPBI C.%
\BCBL {}\ \BBA {} {Mulligan}, P.%
\end{APACrefauthors}%
\unskip\
\newblock
\APACrefYearMonthDay{2015}{{\APACmonth{01}}}{}.
\newblock
{\BBOQ}\APACrefatitle {{Sunjammer}} {{Sunjammer}}.{\BBCQ}
\newblock
\APACjournalVolNumPages{Weather}{70}{1}{27-30}.
\newblock
\begin{APACrefDOI} \doi{10.1002/wea.2438} \end{APACrefDOI}
\PrintBackRefs{\CurrentBib}

\bibitem [\protect \citeauthoryear {%
{Eyles}%
\ \protect \BOthers {.}}{%
{Eyles}%
\ \protect \BOthers {.}}{%
{\protect \APACyear {2009}}%
}]{%
eyles2009heliospheric}
\APACinsertmetastar {%
eyles2009heliospheric}%
\begin{APACrefauthors}%
{Eyles}, C\BPBI J.%
, {Harrison}, R\BPBI A.%
, {Davis}, C\BPBI J.%
, {Waltham}, N\BPBI R.%
, {Shaughnessy}, B\BPBI M.%
, {Mapson-Menard}, H\BPBI C\BPBI A.%
\BDBL {}others%
\end{APACrefauthors}%
\unskip\
\newblock
\APACrefYearMonthDay{2009}{{\APACmonth{02}}}{}.
\newblock
{\BBOQ}\APACrefatitle {{The Heliospheric Imagers Onboard the STEREO Mission}} {{The Heliospheric Imagers Onboard the STEREO Mission}}.{\BBCQ}
\newblock
\APACjournalVolNumPages{\solphys}{254}{2}{387-445}.
\newblock
\begin{APACrefDOI} \doi{10.1007/s11207-008-9299-0} \end{APACrefDOI}
\PrintBackRefs{\CurrentBib}

\bibitem [\protect \citeauthoryear {%
{Good}%
\ \BBA {} {Forsyth}%
}{%
{Good}%
\ \BBA {} {Forsyth}%
}{%
{\protect \APACyear {2016}}%
}]{%
good2016interplanetary}
\APACinsertmetastar {%
good2016interplanetary}%
\begin{APACrefauthors}%
{Good}, S\BPBI W.%
\BCBT {}\ \BBA {} {Forsyth}, R\BPBI J.%
\end{APACrefauthors}%
\unskip\
\newblock
\APACrefYearMonthDay{2016}{{\APACmonth{01}}}{}.
\newblock
{\BBOQ}\APACrefatitle {{Interplanetary Coronal Mass Ejections Observed by MESSENGER and Venus Express}} {{Interplanetary Coronal Mass Ejections Observed by MESSENGER and Venus Express}}.{\BBCQ}
\newblock
\APACjournalVolNumPages{\solphys}{291}{1}{239-263}.
\newblock
\begin{APACrefDOI} \doi{10.1007/s11207-015-0828-3} \end{APACrefDOI}
\PrintBackRefs{\CurrentBib}

\bibitem [\protect \citeauthoryear {%
{Hapgood}%
}{%
{Hapgood}%
}{%
{\protect \APACyear {1992}}%
}]{%
hapgood1992coordinates}
\APACinsertmetastar {%
hapgood1992coordinates}%
\begin{APACrefauthors}%
{Hapgood}, M\BPBI A.%
\end{APACrefauthors}%
\unskip\
\newblock
\APACrefYearMonthDay{1992}{{\APACmonth{05}}}{}.
\newblock
{\BBOQ}\APACrefatitle {{Space physics coordinate transformations: A user guide}} {{Space physics coordinate transformations: A user guide}}.{\BBCQ}
\newblock
\APACjournalVolNumPages{\planss}{40}{5}{711-717}.
\newblock
\begin{APACrefDOI} \doi{10.1016/0032-0633(92)90012-D} \end{APACrefDOI}
\PrintBackRefs{\CurrentBib}

\bibitem [\protect \citeauthoryear {%
{Harvey}%
\ \protect \BOthers {.}}{%
{Harvey}%
\ \protect \BOthers {.}}{%
{\protect \APACyear {1996}}%
}]{%
harvey1996GONG}
\APACinsertmetastar {%
harvey1996GONG}%
\begin{APACrefauthors}%
{Harvey}, J\BPBI W.%
, {Hill}, F.%
, {Hubbard}, R\BPBI P.%
, {Kennedy}, J\BPBI R.%
, {Leibacher}, J\BPBI W.%
, {Pintar}, J\BPBI A.%
\BDBL {}{Yasukawa}, E.%
\end{APACrefauthors}%
\unskip\
\newblock
\APACrefYearMonthDay{1996}{{\APACmonth{05}}}{}.
\newblock
{\BBOQ}\APACrefatitle {{The Global Oscillation Network Group (GONG) Project}} {{The Global Oscillation Network Group (GONG) Project}}.{\BBCQ}
\newblock
\APACjournalVolNumPages{Science}{272}{5266}{1284-1286}.
\newblock
\begin{APACrefDOI} \doi{10.1126/science.272.5266.1284} \end{APACrefDOI}
\PrintBackRefs{\CurrentBib}

\bibitem [\protect \citeauthoryear {%
Hayakawa%
\ \protect \BOthers {.}}{%
Hayakawa%
\ \protect \BOthers {.}}{%
{\protect \APACyear {2024}}%
}]{%
Hayakawa2024}
\APACinsertmetastar {%
Hayakawa2024}%
\begin{APACrefauthors}%
Hayakawa, H.%
, Ebihara, Y.%
, Mishev, A.%
, Koldobskiy, S.%
, Kusano, K.%
, Bechet, S.%
\BDBL {}Miyoshi, Y.%
\end{APACrefauthors}%
\unskip\
\newblock
\APACrefYearMonthDay{2024}{07}{}.
\newblock
\APACrefbtitle {The Solar and Geomagnetic Storms in May 2024: A Flash Data Report.} {The solar and geomagnetic storms in may 2024: A flash data report.}
\newblock
\begin{APACrefDOI} \doi{10.48550/arXiv.2407.07665} \end{APACrefDOI}
\PrintBackRefs{\CurrentBib}

\bibitem [\protect \citeauthoryear {%
{Henon}%
}{%
{Henon}%
}{%
{\protect \APACyear {1969}}%
}]{%
henon1969}
\APACinsertmetastar {%
henon1969}%
\begin{APACrefauthors}%
{Henon}, M.%
\end{APACrefauthors}%
\unskip\
\newblock
\APACrefYearMonthDay{1969}{{\APACmonth{02}}}{}.
\newblock
{\BBOQ}\APACrefatitle {{Numerical exploration of the restricted problem, V}} {{Numerical exploration of the restricted problem, V}}.{\BBCQ}
\newblock
\APACjournalVolNumPages{\aap}{1}{}{223-238}.
\PrintBackRefs{\CurrentBib}

\bibitem [\protect \citeauthoryear {%
{Jarolim}%
, {Veronig}%
, {Purkhart}%
, {Zhang}%
\BCBL {}\ \BBA {} {Rempel}%
}{%
{Jarolim}%
\ \protect \BOthers {.}}{%
{\protect \APACyear {2024}}%
}]{%
Jarolim2024}
\APACinsertmetastar {%
Jarolim2024}%
\begin{APACrefauthors}%
{Jarolim}, R.%
, {Veronig}, A.%
, {Purkhart}, S.%
, {Zhang}, P.%
\BCBL {}\ \BBA {} {Rempel}, M.%
\end{APACrefauthors}%
\unskip\
\newblock
\APACrefYearMonthDay{2024}{{\APACmonth{09}}}{}.
\newblock
{\BBOQ}\APACrefatitle {{Magnetic Field Evolution of the Solar Active Region 13664}} {{Magnetic Field Evolution of the Solar Active Region 13664}}.{\BBCQ}
\newblock
\APACjournalVolNumPages{arXiv e-prints}{}{}{arXiv:2409.08124}.
\newblock
\begin{APACrefDOI} \doi{10.48550/arXiv.2409.08124} \end{APACrefDOI}
\PrintBackRefs{\CurrentBib}

\bibitem [\protect \citeauthoryear {%
Ji%
, Moon%
, Gopalswamy%
\BCBL {}\ \BBA {} Lee%
}{%
Ji%
\ \protect \BOthers {.}}{%
{\protect \APACyear {2012}}%
}]{%
ji2012}
\APACinsertmetastar {%
ji2012}%
\begin{APACrefauthors}%
Ji, E\BHBI Y.%
, Moon, Y\BHBI J.%
, Gopalswamy, N.%
\BCBL {}\ \BBA {} Lee, D\BHBI H.%
\end{APACrefauthors}%
\unskip\
\newblock
\APACrefYearMonthDay{2012}{}{}.
\newblock
{\BBOQ}\APACrefatitle {Comparison of Dst forecast models for intense geomagnetic storms} {Comparison of dst forecast models for intense geomagnetic storms}.{\BBCQ}
\newblock
\APACjournalVolNumPages{Journal of Geophysical Research: Space Physics}{117}{A3}{}.
\newblock
\begin{APACrefURL} \url{https://agupubs.onlinelibrary.wiley.com/doi/abs/10.1029/2011JA016872} \end{APACrefURL}
\newblock
\begin{APACrefDOI} \doi{https://doi.org/10.1029/2011JA016872} \end{APACrefDOI}
\PrintBackRefs{\CurrentBib}

\bibitem [\protect \citeauthoryear {%
{Kaiser}%
\ \protect \BOthers {.}}{%
{Kaiser}%
\ \protect \BOthers {.}}{%
{\protect \APACyear {2008}}%
}]{%
kaiser2008stereo}
\APACinsertmetastar {%
kaiser2008stereo}%
\begin{APACrefauthors}%
{Kaiser}, M\BPBI L.%
, {Kucera}, T\BPBI A.%
, {Davila}, J\BPBI M.%
, {St. Cyr}, O\BPBI C.%
, {Guhathakurta}, M.%
\BCBL {}\ \BBA {} {Christian}, E.%
\end{APACrefauthors}%
\unskip\
\newblock
\APACrefYearMonthDay{2008}{{\APACmonth{04}}}{}.
\newblock
{\BBOQ}\APACrefatitle {{The STEREO Mission: An Introduction}} {{The STEREO Mission: An Introduction}}.{\BBCQ}
\newblock
\APACjournalVolNumPages{\ssr}{136}{1-4}{5-16}.
\newblock
\begin{APACrefDOI} \doi{10.1007/s11214-007-9277-0} \end{APACrefDOI}
\PrintBackRefs{\CurrentBib}

\bibitem [\protect \citeauthoryear {%
{Kay}%
\ \protect \BOthers {.}}{%
{Kay}%
\ \protect \BOthers {.}}{%
{\protect \APACyear {2024}}%
}]{%
kay2024CMEarrivals}
\APACinsertmetastar {%
kay2024CMEarrivals}%
\begin{APACrefauthors}%
{Kay}, C.%
, {Palmerio}, E.%
, {Riley}, P.%
, {Mays}, M\BPBI L.%
, {Nieves-Chinchilla}, T.%
, {Romano}, M.%
\BDBL {}{Chulaki}, A.%
\end{APACrefauthors}%
\unskip\
\newblock
\APACrefYearMonthDay{2024}{{\APACmonth{07}}}{}.
\newblock
{\BBOQ}\APACrefatitle {{Updating Measures of CME Arrival Time Errors}} {{Updating Measures of CME Arrival Time Errors}}.{\BBCQ}
\newblock
\APACjournalVolNumPages{Space Weather}{22}{7}{e2024SW003951}.
\newblock
\begin{APACrefDOI} \doi{10.1029/2024SW003951} \end{APACrefDOI}
\PrintBackRefs{\CurrentBib}

\bibitem [\protect \citeauthoryear {%
Kilpua%
, Lugaz%
, Mays%
\BCBL {}\ \BBA {} Temmer%
}{%
Kilpua%
\ \protect \BOthers {.}}{%
{\protect \APACyear {2019}}%
}]{%
kilpua2019review}
\APACinsertmetastar {%
kilpua2019review}%
\begin{APACrefauthors}%
Kilpua, E.%
, Lugaz, N.%
, Mays, M\BPBI L.%
\BCBL {}\ \BBA {} Temmer, M.%
\end{APACrefauthors}%
\unskip\
\newblock
\APACrefYearMonthDay{2019}{}{}.
\newblock
{\BBOQ}\APACrefatitle {Forecasting the Structure and Orientation of Earthbound Coronal Mass Ejections} {Forecasting the structure and orientation of earthbound coronal mass ejections}.{\BBCQ}
\newblock
\APACjournalVolNumPages{Space Weather}{17}{4}{498-526}.
\newblock
\begin{APACrefURL} \url{https://agupubs.onlinelibrary.wiley.com/doi/abs/10.1029/2018SW001944} \end{APACrefURL}
\newblock
\begin{APACrefDOI} \doi{https://doi.org/10.1029/2018SW001944} \end{APACrefDOI}
\PrintBackRefs{\CurrentBib}

\bibitem [\protect \citeauthoryear {%
Koehn%
\ \protect \BOthers {.}}{%
Koehn%
\ \protect \BOthers {.}}{%
{\protect \APACyear {2022}}%
}]{%
koehn2022}
\APACinsertmetastar {%
koehn2022}%
\begin{APACrefauthors}%
Koehn, G\BPBI J.%
, Desai, R\BPBI T.%
, Davies, E\BPBI E.%
, Forsyth, R\BPBI J.%
, Eastwood, J\BPBI P.%
\BCBL {}\ \BBA {} Poedts, S.%
\end{APACrefauthors}%
\unskip\
\newblock
\APACrefYearMonthDay{2022}{dec}{}.
\newblock
{\BBOQ}\APACrefatitle {Successive Interacting Coronal Mass Ejections: How to Create a Perfect Storm} {Successive interacting coronal mass ejections: How to create a perfect storm}.{\BBCQ}
\newblock
\APACjournalVolNumPages{The Astrophysical Journal}{941}{2}{139}.
\newblock
\begin{APACrefURL} \url{https://dx.doi.org/10.3847/1538-4357/aca28c} \end{APACrefURL}
\newblock
\begin{APACrefDOI} \doi{10.3847/1538-4357/aca28c} \end{APACrefDOI}
\PrintBackRefs{\CurrentBib}

\bibitem [\protect \citeauthoryear {%
{Kubicka}%
\ \protect \BOthers {.}}{%
{Kubicka}%
\ \protect \BOthers {.}}{%
{\protect \APACyear {2016}}%
}]{%
kubicka2016dst}
\APACinsertmetastar {%
kubicka2016dst}%
\begin{APACrefauthors}%
{Kubicka}, M.%
, {M{\"o}stl}, C.%
, {Amerstorfer}, T.%
, {Boakes}, P\BPBI D.%
, {Feng}, L.%
, {Eastwood}, J\BPBI P.%
\BCBL {}\ \BBA {} {T{\"o}rm{\"a}nen}, O.%
\end{APACrefauthors}%
\unskip\
\newblock
\APACrefYearMonthDay{2016}{{\APACmonth{12}}}{}.
\newblock
{\BBOQ}\APACrefatitle {{Prediction of Geomagnetic Storm Strength from Inner Heliospheric In Situ Observations}} {{Prediction of Geomagnetic Storm Strength from Inner Heliospheric In Situ Observations}}.{\BBCQ}
\newblock
\APACjournalVolNumPages{\apj}{833}{2}{255}.
\newblock
\begin{APACrefDOI} \doi{10.3847/1538-4357/833/2/255} \end{APACrefDOI}
\PrintBackRefs{\CurrentBib}

\bibitem [\protect \citeauthoryear {%
{Laker}%
\ \protect \BOthers {.}}{%
{Laker}%
\ \protect \BOthers {.}}{%
{\protect \APACyear {2024}}%
}]{%
laker_2024}
\APACinsertmetastar {%
laker_2024}%
\begin{APACrefauthors}%
{Laker}, R.%
, {Horbury}, T\BPBI S.%
, {O'Brien}, H.%
, {Fauchon-Jones}, E\BPBI J.%
, {Angelini}, V.%
, {Fargette}, N.%
\BDBL {}{Dumbovi{\'c}}, M.%
\end{APACrefauthors}%
\unskip\
\newblock
\APACrefYearMonthDay{2024}{{\APACmonth{02}}}{}.
\newblock
{\BBOQ}\APACrefatitle {{Using Solar Orbiter as an Upstream Solar Wind Monitor for Real Time Space Weather Predictions}} {{Using Solar Orbiter as an Upstream Solar Wind Monitor for Real Time Space Weather Predictions}}.{\BBCQ}
\newblock
\APACjournalVolNumPages{Space Weather}{22}{2}{e2023SW003628}.
\newblock
\begin{APACrefDOI} \doi{10.1029/2023SW003628} \end{APACrefDOI}
\PrintBackRefs{\CurrentBib}

\bibitem [\protect \citeauthoryear {%
{Lemen}%
\ \protect \BOthers {.}}{%
{Lemen}%
\ \protect \BOthers {.}}{%
{\protect \APACyear {2012}}%
}]{%
lemen2011atmospheric}
\APACinsertmetastar {%
lemen2011atmospheric}%
\begin{APACrefauthors}%
{Lemen}, J\BPBI R.%
, {Title}, A\BPBI M.%
, {Akin}, D\BPBI J.%
, {Boerner}, P\BPBI F.%
, {Chou}, C.%
, {Drake}, J\BPBI F.%
\BDBL {}others%
\end{APACrefauthors}%
\unskip\
\newblock
\APACrefYearMonthDay{2012}{{\APACmonth{01}}}{}.
\newblock
{\BBOQ}\APACrefatitle {{The Atmospheric Imaging Assembly (AIA) on the Solar Dynamics Observatory (SDO)}} {{The Atmospheric Imaging Assembly (AIA) on the Solar Dynamics Observatory (SDO)}}.{\BBCQ}
\newblock
\APACjournalVolNumPages{\solphys}{275}{1-2}{17-40}.
\newblock
\begin{APACrefDOI} \doi{10.1007/s11207-011-9776-8} \end{APACrefDOI}
\PrintBackRefs{\CurrentBib}

\bibitem [\protect \citeauthoryear {%
{Lindsay}%
, {Russell}%
\BCBL {}\ \BBA {} {Luhmann}%
}{%
{Lindsay}%
\ \protect \BOthers {.}}{%
{\protect \APACyear {1999}}%
}]{%
lindsay1999dst}
\APACinsertmetastar {%
lindsay1999dst}%
\begin{APACrefauthors}%
{Lindsay}, G\BPBI M.%
, {Russell}, C\BPBI T.%
\BCBL {}\ \BBA {} {Luhmann}, J\BPBI G.%
\end{APACrefauthors}%
\unskip\
\newblock
\APACrefYearMonthDay{1999}{{\APACmonth{05}}}{}.
\newblock
{\BBOQ}\APACrefatitle {{Predictability of Dst index based upon solar wind conditions monitored inside 1 AU}} {{Predictability of Dst index based upon solar wind conditions monitored inside 1 AU}}.{\BBCQ}
\newblock
\APACjournalVolNumPages{\jgr}{104}{A5}{10335-10344}.
\newblock
\begin{APACrefDOI} \doi{10.1029/1999JA900010} \end{APACrefDOI}
\PrintBackRefs{\CurrentBib}

\bibitem [\protect \citeauthoryear {%
Liu%
, Hu%
, Zhao%
, Chen%
\BCBL {}\ \BBA {} Wang%
}{%
Liu%
\ \protect \BOthers {.}}{%
{\protect \APACyear {2024}}%
}]{%
liu2024may}
\APACinsertmetastar {%
liu2024may}%
\begin{APACrefauthors}%
Liu, Y\BPBI D.%
, Hu, H.%
, Zhao, X.%
, Chen, C.%
\BCBL {}\ \BBA {} Wang, R.%
\end{APACrefauthors}%
\unskip\
\newblock
\APACrefYearMonthDay{2024}{}{}.
\newblock
\APACrefbtitle {A Pileup of Coronal Mass Ejections Produced the Largest Geomagnetic Storm in Two Decades.} {A pileup of coronal mass ejections produced the largest geomagnetic storm in two decades.}
\newblock
\begin{APACrefURL} \url{https://arxiv.org/abs/2409.11492} \end{APACrefURL}
\PrintBackRefs{\CurrentBib}

\bibitem [\protect \citeauthoryear {%
{Liu}%
\ \protect \BOthers {.}}{%
{Liu}%
\ \protect \BOthers {.}}{%
{\protect \APACyear {2014}}%
}]{%
liu2014observations}
\APACinsertmetastar {%
liu2014observations}%
\begin{APACrefauthors}%
{Liu}, Y\BPBI D.%
, {Luhmann}, J\BPBI G.%
, {Kajdi{\v{c}}}, P.%
, {Kilpua}, E\BPBI K\BPBI J.%
, {Lugaz}, N.%
, {Nitta}, N\BPBI V.%
\BDBL {}others%
\end{APACrefauthors}%
\unskip\
\newblock
\APACrefYearMonthDay{2014}{{\APACmonth{03}}}{}.
\newblock
{\BBOQ}\APACrefatitle {{Observations of an extreme storm in interplanetary space caused by successive coronal mass ejections}} {{Observations of an extreme storm in interplanetary space caused by successive coronal mass ejections}}.{\BBCQ}
\newblock
\APACjournalVolNumPages{\natcom}{5}{}{3481}.
\newblock
\begin{APACrefDOI} \doi{10.1038/ncomms4481} \end{APACrefDOI}
\PrintBackRefs{\CurrentBib}

\bibitem [\protect \citeauthoryear {%
{Love}%
, {Rigler}%
, {Hayakawa}%
\BCBL {}\ \BBA {} {Mursula}%
}{%
{Love}%
\ \protect \BOthers {.}}{%
{\protect \APACyear {2024}}%
}]{%
love2024carrington}
\APACinsertmetastar {%
love2024carrington}%
\begin{APACrefauthors}%
{Love}, J\BPBI J.%
, {Rigler}, E\BPBI J.%
, {Hayakawa}, H.%
\BCBL {}\ \BBA {} {Mursula}, K.%
\end{APACrefauthors}%
\unskip\
\newblock
\APACrefYearMonthDay{2024}{{\APACmonth{08}}}{}.
\newblock
{\BBOQ}\APACrefatitle {{On the uncertain intensity estimate of the 1859 Carrington storm}} {{On the uncertain intensity estimate of the 1859 Carrington storm}}.{\BBCQ}
\newblock
\APACjournalVolNumPages{Journal of Space Weather and Space Climate}{14}{}{21}.
\newblock
\begin{APACrefDOI} \doi{10.1051/swsc/2024015} \end{APACrefDOI}
\PrintBackRefs{\CurrentBib}

\bibitem [\protect \citeauthoryear {%
{Love}%
, {Rigler}%
, {Pulkkinen}%
\BCBL {}\ \BBA {} {Riley}%
}{%
{Love}%
\ \protect \BOthers {.}}{%
{\protect \APACyear {2015}}%
}]{%
love2015_historic}
\APACinsertmetastar {%
love2015_historic}%
\begin{APACrefauthors}%
{Love}, J\BPBI J.%
, {Rigler}, E\BPBI J.%
, {Pulkkinen}, A.%
\BCBL {}\ \BBA {} {Riley}, P.%
\end{APACrefauthors}%
\unskip\
\newblock
\APACrefYearMonthDay{2015}{{\APACmonth{08}}}{}.
\newblock
{\BBOQ}\APACrefatitle {{On the lognormality of historical magnetic storm intensity statistics: Implications for extreme-event probabilities}} {{On the lognormality of historical magnetic storm intensity statistics: Implications for extreme-event probabilities}}.{\BBCQ}
\newblock
\APACjournalVolNumPages{\grl}{42}{16}{6544-6553}.
\newblock
\begin{APACrefDOI} \doi{10.1002/2015GL064842} \end{APACrefDOI}
\PrintBackRefs{\CurrentBib}

\bibitem [\protect \citeauthoryear {%
Lugaz%
\ \protect \BOthers {.}}{%
Lugaz%
\ \protect \BOthers {.}}{%
{\protect \APACyear {2016}}%
}]{%
lugaz2016shocks}
\APACinsertmetastar {%
lugaz2016shocks}%
\begin{APACrefauthors}%
Lugaz, N.%
, Farrugia, C\BPBI J.%
, Winslow, R\BPBI M.%
, Al-Haddad, N.%
, Kilpua, E\BPBI K\BPBI J.%
\BCBL {}\ \BBA {} Riley, P.%
\end{APACrefauthors}%
\unskip\
\newblock
\APACrefYearMonthDay{2016}{}{}.
\newblock
{\BBOQ}\APACrefatitle {Factors affecting the geoeffectiveness of shocks and sheaths at 1 AU} {Factors affecting the geoeffectiveness of shocks and sheaths at 1 au}.{\BBCQ}
\newblock
\APACjournalVolNumPages{Journal of Geophysical Research: Space Physics}{121}{11}{10,861-10,879}.
\newblock
\begin{APACrefURL} \url{https://agupubs.onlinelibrary.wiley.com/doi/abs/10.1002/2016JA023100} \end{APACrefURL}
\newblock
\begin{APACrefDOI} \doi{https://doi.org/10.1002/2016JA023100} \end{APACrefDOI}
\PrintBackRefs{\CurrentBib}

\bibitem [\protect \citeauthoryear {%
Lugaz%
, IV%
\BCBL {}\ \BBA {} Gombosi%
}{%
Lugaz%
\ \protect \BOthers {.}}{%
{\protect \APACyear {2005}}%
}]{%
lugaz2005b}
\APACinsertmetastar {%
lugaz2005b}%
\begin{APACrefauthors}%
Lugaz, N.%
, IV, W\BPBI B\BPBI M.%
\BCBL {}\ \BBA {} Gombosi, T\BPBI I.%
\end{APACrefauthors}%
\unskip\
\newblock
\APACrefYearMonthDay{2005}{nov}{}.
\newblock
{\BBOQ}\APACrefatitle {Numerical Simulation of the Interaction of Two Coronal Mass Ejections from Sun to Earth} {Numerical simulation of the interaction of two coronal mass ejections from sun to earth}.{\BBCQ}
\newblock
\APACjournalVolNumPages{The Astrophysical Journal}{634}{1}{651}.
\newblock
\begin{APACrefURL} \url{https://dx.doi.org/10.1086/491782} \end{APACrefURL}
\newblock
\begin{APACrefDOI} \doi{10.1086/491782} \end{APACrefDOI}
\PrintBackRefs{\CurrentBib}

\bibitem [\protect \citeauthoryear {%
Lugaz%
, Lee%
\BCBL {}\ \protect \BOthers {.}}{%
Lugaz%
, Lee%
\BCBL {}\ \protect \BOthers {.}}{%
{\protect \APACyear {2024}}%
}]{%
lugaz2024mission}
\APACinsertmetastar {%
lugaz2024mission}%
\begin{APACrefauthors}%
Lugaz, N.%
, Lee, C\BPBI O.%
, Al-Haddad, N.%
, Lillis, R\BPBI J.%
, Jian, L\BPBI K.%
, Curtis, D\BPBI W.%
\BDBL {}Nieves-Chinchilla, T.%
\end{APACrefauthors}%
\unskip\
\newblock
\APACrefYearMonthDay{2024}{Sep}{20}.
\newblock
{\BBOQ}\APACrefatitle {The Need for Near-Earth Multi-Spacecraft Heliospheric Measurements and an Explorer Mission to Investigate Interplanetary Structures and Transients in the Near-Earth Heliosphere} {The need for near-earth multi-spacecraft heliospheric measurements and an explorer mission to investigate interplanetary structures and transients in the near-earth heliosphere}.{\BBCQ}
\newblock
\APACjournalVolNumPages{Space Science Reviews}{220}{7}{73}.
\newblock
\begin{APACrefURL} \url{https://doi.org/10.1007/s11214-024-01108-8} \end{APACrefURL}
\newblock
\begin{APACrefDOI} \doi{10.1007/s11214-024-01108-8} \end{APACrefDOI}
\PrintBackRefs{\CurrentBib}

\bibitem [\protect \citeauthoryear {%
Lugaz%
, Zhuang%
\BCBL {}\ \protect \BOthers {.}}{%
Lugaz%
, Zhuang%
\BCBL {}\ \protect \BOthers {.}}{%
{\protect \APACyear {2024}}%
}]{%
lugaz2024MEwidth}
\APACinsertmetastar {%
lugaz2024MEwidth}%
\begin{APACrefauthors}%
Lugaz, N.%
, Zhuang, B.%
, Scolini, C.%
, Al-Haddad, N.%
, Farrugia, C\BPBI J.%
, Winslow, R\BPBI M.%
\BDBL {}Galvin, A\BPBI B.%
\end{APACrefauthors}%
\unskip\
\newblock
\APACrefYearMonthDay{2024}{feb}{}.
\newblock
{\BBOQ}\APACrefatitle {The Width of Magnetic Ejecta Measured near 1 au: Lessons from STEREO-A Measurements in 2021–2022} {The width of magnetic ejecta measured near 1 au: Lessons from stereo-a measurements in 2021–2022}.{\BBCQ}
\newblock
\APACjournalVolNumPages{The Astrophysical Journal}{962}{2}{193}.
\newblock
\begin{APACrefURL} \url{https://dx.doi.org/10.3847/1538-4357/ad17b9} \end{APACrefURL}
\newblock
\begin{APACrefDOI} \doi{10.3847/1538-4357/ad17b9} \end{APACrefDOI}
\PrintBackRefs{\CurrentBib}

\bibitem [\protect \citeauthoryear {%
Lundstedt%
, Gleisner%
\BCBL {}\ \BBA {} Wintoft%
}{%
Lundstedt%
\ \protect \BOthers {.}}{%
{\protect \APACyear {2002}}%
}]{%
lundstedt2002}
\APACinsertmetastar {%
lundstedt2002}%
\begin{APACrefauthors}%
Lundstedt, H.%
, Gleisner, H.%
\BCBL {}\ \BBA {} Wintoft, P.%
\end{APACrefauthors}%
\unskip\
\newblock
\APACrefYearMonthDay{2002}{}{}.
\newblock
{\BBOQ}\APACrefatitle {Operational forecasts of the geomagnetic Dst index} {Operational forecasts of the geomagnetic dst index}.{\BBCQ}
\newblock
\APACjournalVolNumPages{Geophysical Research Letters}{29}{24}{34-1-34-4}.
\newblock
\begin{APACrefURL} \url{https://agupubs.onlinelibrary.wiley.com/doi/abs/10.1029/2002GL016151} \end{APACrefURL}
\newblock
\begin{APACrefDOI} \doi{https://doi.org/10.1029/2002GL016151} \end{APACrefDOI}
\PrintBackRefs{\CurrentBib}

\bibitem [\protect \citeauthoryear {%
Mac~Manus%
\ \protect \BOthers {.}}{%
Mac~Manus%
\ \protect \BOthers {.}}{%
{\protect \APACyear {2023}}%
}]{%
manus2023NZmitigation}
\APACinsertmetastar {%
manus2023NZmitigation}%
\begin{APACrefauthors}%
Mac~Manus, D\BPBI H.%
, Rodger, C\BPBI J.%
, Renton, A.%
, Ronald, J.%
, Harper, D.%
, Taylor, C.%
\BDBL {}Clilverd, M\BPBI A.%
\end{APACrefauthors}%
\unskip\
\newblock
\APACrefYearMonthDay{2023}{}{}.
\newblock
{\BBOQ}\APACrefatitle {Geomagnetically Induced Current Mitigation in New Zealand: Operational Mitigation Method Development With Industry Input} {Geomagnetically induced current mitigation in new zealand: Operational mitigation method development with industry input}.{\BBCQ}
\newblock
\APACjournalVolNumPages{Space Weather}{21}{11}{e2023SW003533}.
\newblock
\begin{APACrefURL} \url{https://agupubs.onlinelibrary.wiley.com/doi/abs/10.1029/2023SW003533} \end{APACrefURL}
\newblock
\APACrefnote{e2023SW003533 2023SW003533}
\newblock
\begin{APACrefDOI} \doi{https://doi.org/10.1029/2023SW003533} \end{APACrefDOI}
\PrintBackRefs{\CurrentBib}

\bibitem [\protect \citeauthoryear {%
{Martin}%
}{%
{Martin}%
}{%
{\protect \APACyear {1998}}%
}]{%
martin1998}
\APACinsertmetastar {%
martin1998}%
\begin{APACrefauthors}%
{Martin}, S\BPBI F.%
\end{APACrefauthors}%
\unskip\
\newblock
\APACrefYearMonthDay{1998}{{\APACmonth{09}}}{}.
\newblock
{\BBOQ}\APACrefatitle {{Conditions for the Formation and Maintenance of Filaments (Invited Review)}} {{Conditions for the Formation and Maintenance of Filaments (Invited Review)}}.{\BBCQ}
\newblock
\APACjournalVolNumPages{\solphys}{182}{1}{107-137}.
\newblock
\begin{APACrefDOI} \doi{10.1023/A:1005026814076} \end{APACrefDOI}
\PrintBackRefs{\CurrentBib}

\bibitem [\protect \citeauthoryear {%
{McComas}%
\ \protect \BOthers {.}}{%
{McComas}%
\ \protect \BOthers {.}}{%
{\protect \APACyear {1998}}%
}]{%
mccomas1998solar}
\APACinsertmetastar {%
mccomas1998solar}%
\begin{APACrefauthors}%
{McComas}, D\BPBI J.%
, {Bame}, S\BPBI J.%
, {Barker}, P.%
, {Feldman}, W\BPBI C.%
, {Phillips}, J\BPBI L.%
, {Riley}, P.%
\BCBL {}\ \BBA {} {Griffee}, J\BPBI W.%
\end{APACrefauthors}%
\unskip\
\newblock
\APACrefYearMonthDay{1998}{{\APACmonth{07}}}{}.
\newblock
{\BBOQ}\APACrefatitle {{Solar Wind Electron Proton Alpha Monitor (SWEPAM) for the Advanced Composition Explorer}} {{Solar Wind Electron Proton Alpha Monitor (SWEPAM) for the Advanced Composition Explorer}}.{\BBCQ}
\newblock
\APACjournalVolNumPages{\ssr}{86}{}{563-612}.
\newblock
\begin{APACrefDOI} \doi{10.1023/A:1005040232597} \end{APACrefDOI}
\PrintBackRefs{\CurrentBib}

\bibitem [\protect \citeauthoryear {%
{Meng}%
, {Tsurutani}%
\BCBL {}\ \BBA {} {Mannucci}%
}{%
{Meng}%
\ \protect \BOthers {.}}{%
{\protect \APACyear {2019}}%
}]{%
meng2019superstorms}
\APACinsertmetastar {%
meng2019superstorms}%
\begin{APACrefauthors}%
{Meng}, X.%
, {Tsurutani}, B\BPBI T.%
\BCBL {}\ \BBA {} {Mannucci}, A\BPBI J.%
\end{APACrefauthors}%
\unskip\
\newblock
\APACrefYearMonthDay{2019}{{\APACmonth{06}}}{}.
\newblock
{\BBOQ}\APACrefatitle {{The Solar and Interplanetary Causes of Superstorms (Minimum Dst {\ensuremath{\leq}} -250 nT) During the Space Age}} {{The Solar and Interplanetary Causes of Superstorms (Minimum Dst {\ensuremath{\leq}} -250 nT) During the Space Age}}.{\BBCQ}
\newblock
\APACjournalVolNumPages{Journal of Geophysical Research (Space Physics)}{124}{6}{3926-3948}.
\newblock
\begin{APACrefDOI} \doi{10.1029/2018JA026425} \end{APACrefDOI}
\PrintBackRefs{\CurrentBib}

\bibitem [\protect \citeauthoryear {%
Morley%
}{%
Morley%
}{%
{\protect \APACyear {2020}}%
}]{%
morley2020}
\APACinsertmetastar {%
morley2020}%
\begin{APACrefauthors}%
Morley, S\BPBI K.%
\end{APACrefauthors}%
\unskip\
\newblock
\APACrefYearMonthDay{2020}{}{}.
\newblock
{\BBOQ}\APACrefatitle {Challenges and Opportunities in Magnetospheric Space Weather Prediction} {Challenges and opportunities in magnetospheric space weather prediction}.{\BBCQ}
\newblock
\APACjournalVolNumPages{Space Weather}{18}{3}{e2018SW002108}.
\newblock
\begin{APACrefURL} \url{https://agupubs.onlinelibrary.wiley.com/doi/abs/10.1029/2018SW002108} \end{APACrefURL}
\newblock
\APACrefnote{e2018SW002108 10.1029/2018SW002108}
\newblock
\begin{APACrefDOI} \doi{https://doi.org/10.1029/2018SW002108} \end{APACrefDOI}
\PrintBackRefs{\CurrentBib}

\bibitem [\protect \citeauthoryear {%
{M{\"o}stl}%
\ \protect \BOthers {.}}{%
{M{\"o}stl}%
\ \protect \BOthers {.}}{%
{\protect \APACyear {2015}}%
}]{%
moestl2015elevo}
\APACinsertmetastar {%
moestl2015elevo}%
\begin{APACrefauthors}%
{M{\"o}stl}, C.%
, {Rollett}, T.%
, {Frahm}, R\BPBI A.%
, {Liu}, Y\BPBI D.%
, {Long}, D\BPBI M.%
, {Colaninno}, R\BPBI C.%
\BDBL {}{Vr{\v{s}}nak}, B.%
\end{APACrefauthors}%
\unskip\
\newblock
\APACrefYearMonthDay{2015}{{\APACmonth{05}}}{}.
\newblock
{\BBOQ}\APACrefatitle {{Strong coronal channelling and interplanetary evolution of a solar storm up to Earth and Mars}} {{Strong coronal channelling and interplanetary evolution of a solar storm up to Earth and Mars}}.{\BBCQ}
\newblock
\APACjournalVolNumPages{Nature Communications}{6}{}{7135}.
\newblock
\begin{APACrefDOI} \doi{10.1038/ncomms8135} \end{APACrefDOI}
\PrintBackRefs{\CurrentBib}

\bibitem [\protect \citeauthoryear {%
M\"ostl%
, Weiss%
, Bailey%
\BCBL {}\ \BBA {} Reiss%
}{%
M\"ostl%
\ \protect \BOthers {.}}{%
{\protect \APACyear {2020}}%
}]{%
moestl2020ICMECAT}
\APACinsertmetastar {%
moestl2020ICMECAT}%
\begin{APACrefauthors}%
M\"ostl, C.%
, Weiss, A.%
, Bailey, R.%
\BCBL {}\ \BBA {} Reiss, M.%
\end{APACrefauthors}%
\unskip\
\newblock
\APACrefYearMonthDay{2020}{Jun}{}.
\newblock
\APACrefbtitle {HELCATS Interplanetary Coronal Mass Ejection Catalog v2.0.} {Helcats interplanetary coronal mass ejection catalog v2.0.}
\newblock
\APACaddressPublisher{}{figshare}.
\newblock
\begin{APACrefURL} \url{https://figshare.com/articles/dataset/HELCATS_Interplanetary_Coronal_Mass_Ejection_Catalog_v2_0/6356420/7} \end{APACrefURL}
\newblock
\begin{APACrefDOI} \doi{10.6084/m9.figshare.6356420.v7} \end{APACrefDOI}
\PrintBackRefs{\CurrentBib}

\bibitem [\protect \citeauthoryear {%
{Nose}%
, {Iyemori}%
, {Sugiura}%
\BCBL {}\ \BBA {} {Kamei}%
}{%
{Nose}%
\ \protect \BOthers {.}}{%
{\protect \APACyear {2015}}%
}]{%
kyoto_dst}
\APACinsertmetastar {%
kyoto_dst}%
\begin{APACrefauthors}%
{Nose}, M.%
, {Iyemori}, T.%
, {Sugiura}, M.%
\BCBL {}\ \BBA {} {Kamei}, T.%
\end{APACrefauthors}%
\unskip\
\newblock
\APACrefYearMonthDay{2015}{}{}.
\newblock
\APACrefbtitle {Geomagnetic Dst index.} {Geomagnetic dst index.}
\newblock
\APACaddressPublisher{}{World Data Center for Geomagnetism, Kyoto}.
\newblock
\begin{APACrefURL} \url{https://isds-datadoi.nict.go.jp/wds/10.17593__14515-74000.html} \end{APACrefURL}
\newblock
\begin{APACrefDOI} \doi{10.17593/14515-74000} \end{APACrefDOI}
\PrintBackRefs{\CurrentBib}

\bibitem [\protect \citeauthoryear {%
O'Brien%
\ \BBA {} McPherron%
}{%
O'Brien%
\ \BBA {} McPherron%
}{%
{\protect \APACyear {2000}}%
}]{%
obrien2000}
\APACinsertmetastar {%
obrien2000}%
\begin{APACrefauthors}%
O'Brien, T.%
\BCBT {}\ \BBA {} McPherron, R\BPBI L.%
\end{APACrefauthors}%
\unskip\
\newblock
\APACrefYearMonthDay{2000}{}{}.
\newblock
{\BBOQ}\APACrefatitle {Forecasting the ring current index Dst in real time} {Forecasting the ring current index dst in real time}.{\BBCQ}
\newblock
\APACjournalVolNumPages{Journal of Atmospheric and Solar-Terrestrial Physics}{62}{14}{1295-1299}.
\newblock
\begin{APACrefURL} \url{https://www.sciencedirect.com/science/article/pii/S1364682600000729} \end{APACrefURL}
\newblock
\APACrefnote{Space Weather Week}
\newblock
\begin{APACrefDOI} \doi{https://doi.org/10.1016/S1364-6826(00)00072-9} \end{APACrefDOI}
\PrintBackRefs{\CurrentBib}

\bibitem [\protect \citeauthoryear {%
{Palmerio}%
\ \protect \BOthers {.}}{%
{Palmerio}%
\ \protect \BOthers {.}}{%
{\protect \APACyear {2017}}%
}]{%
palmerio2017determining}
\APACinsertmetastar {%
palmerio2017determining}%
\begin{APACrefauthors}%
{Palmerio}, E.%
, {Kilpua}, E\BPBI K\BPBI J.%
, {James}, A\BPBI W.%
, {Green}, L\BPBI M.%
, {Pomoell}, J.%
, {Isavnin}, A.%
\BCBL {}\ \BBA {} {Valori}, G.%
\end{APACrefauthors}%
\unskip\
\newblock
\APACrefYearMonthDay{2017}{{\APACmonth{02}}}{}.
\newblock
{\BBOQ}\APACrefatitle {{Determining the Intrinsic CME Flux Rope Type Using Remote-sensing Solar Disk Observations}} {{Determining the Intrinsic CME Flux Rope Type Using Remote-sensing Solar Disk Observations}}.{\BBCQ}
\newblock
\APACjournalVolNumPages{\solphys}{292}{2}{39}.
\newblock
\begin{APACrefDOI} \doi{10.1007/s11207-017-1063-x} \end{APACrefDOI}
\PrintBackRefs{\CurrentBib}

\bibitem [\protect \citeauthoryear {%
{Palmerio}%
\ \protect \BOthers {.}}{%
{Palmerio}%
\ \protect \BOthers {.}}{%
{\protect \APACyear {2018}}%
}]{%
palmerio2018coronal}
\APACinsertmetastar {%
palmerio2018coronal}%
\begin{APACrefauthors}%
{Palmerio}, E.%
, {Kilpua}, E\BPBI K\BPBI J.%
, {M{\"o}stl}, C.%
, {Bothmer}, V.%
, {James}, A\BPBI W.%
, {Green}, L\BPBI M.%
\BDBL {}{Harrison}, R\BPBI A.%
\end{APACrefauthors}%
\unskip\
\newblock
\APACrefYearMonthDay{2018}{{\APACmonth{05}}}{}.
\newblock
{\BBOQ}\APACrefatitle {{Coronal Magnetic Structure of Earthbound CMEs and In Situ Comparison}} {{Coronal Magnetic Structure of Earthbound CMEs and In Situ Comparison}}.{\BBCQ}
\newblock
\APACjournalVolNumPages{Space Weather}{16}{5}{442-460}.
\newblock
\begin{APACrefDOI} \doi{10.1002/2017SW001767} \end{APACrefDOI}
\PrintBackRefs{\CurrentBib}

\bibitem [\protect \citeauthoryear {%
Papitashvili%
\ \BBA {} King%
}{%
Papitashvili%
\ \BBA {} King%
}{%
{\protect \APACyear {2020}}%
{\protect \APACexlab {{\protect \BCnt {1}}}}}]{%
omnimin}
\APACinsertmetastar {%
omnimin}%
\begin{APACrefauthors}%
Papitashvili, N\BPBI E.%
\BCBT {}\ \BBA {} King, J\BPBI H.%
\end{APACrefauthors}%
\unskip\
\newblock
\APACrefYearMonthDay{2020{\protect \BCnt {1}}}{}{}.
\newblock
\APACrefbtitle {OMNI 1-min Data.} {Omni 1-min data.}
\newblock
\APACaddressPublisher{}{NASA Space Physics Data Facility}.
\newblock
\begin{APACrefURL} \url{https://spdf.gsfc.nasa.gov/pub/data/omni/high_res_omni/} \end{APACrefURL}
\newblock
\begin{APACrefDOI} \doi{10.48322/45bb-8792} \end{APACrefDOI}
\PrintBackRefs{\CurrentBib}

\bibitem [\protect \citeauthoryear {%
Papitashvili%
\ \BBA {} King%
}{%
Papitashvili%
\ \BBA {} King%
}{%
{\protect \APACyear {2020}}%
{\protect \APACexlab {{\protect \BCnt {2}}}}}]{%
omnihour}
\APACinsertmetastar {%
omnihour}%
\begin{APACrefauthors}%
Papitashvili, N\BPBI E.%
\BCBT {}\ \BBA {} King, J\BPBI H.%
\end{APACrefauthors}%
\unskip\
\newblock
\APACrefYearMonthDay{2020{\protect \BCnt {2}}}{}{}.
\newblock
\APACrefbtitle {OMNI Hourly Data.} {Omni hourly data.}
\newblock
\APACaddressPublisher{}{NASA Space Physics Data Facility}.
\newblock
\begin{APACrefURL} \url{https://spdf.gsfc.nasa.gov/pub/data/omni/low_res_omni/} \end{APACrefURL}
\newblock
\begin{APACrefDOI} \doi{10.48322/1shr-ht18} \end{APACrefDOI}
\PrintBackRefs{\CurrentBib}

\bibitem [\protect \citeauthoryear {%
{Pesnell}%
, {Thompson}%
\BCBL {}\ \BBA {} {Chamberlin}%
}{%
{Pesnell}%
\ \protect \BOthers {.}}{%
{\protect \APACyear {2012}}%
}]{%
pesnell2012sdo}
\APACinsertmetastar {%
pesnell2012sdo}%
\begin{APACrefauthors}%
{Pesnell}, W\BPBI D.%
, {Thompson}, B\BPBI J.%
\BCBL {}\ \BBA {} {Chamberlin}, P\BPBI C.%
\end{APACrefauthors}%
\unskip\
\newblock
\APACrefYearMonthDay{2012}{{\APACmonth{01}}}{}.
\newblock
{\BBOQ}\APACrefatitle {{The Solar Dynamics Observatory (SDO)}} {{The Solar Dynamics Observatory (SDO)}}.{\BBCQ}
\newblock
\APACjournalVolNumPages{\solphys}{275}{1-2}{3-15}.
\newblock
\begin{APACrefDOI} \doi{10.1007/s11207-011-9841-3} \end{APACrefDOI}
\PrintBackRefs{\CurrentBib}

\bibitem [\protect \citeauthoryear {%
{Pevtsov}%
\ \BBA {} {Balasubramaniam}%
}{%
{Pevtsov}%
\ \BBA {} {Balasubramaniam}%
}{%
{\protect \APACyear {2003}}%
}]{%
pevtsov2003helicityrule}
\APACinsertmetastar {%
pevtsov2003helicityrule}%
\begin{APACrefauthors}%
{Pevtsov}, A\BPBI A.%
\BCBT {}\ \BBA {} {Balasubramaniam}, K\BPBI S.%
\end{APACrefauthors}%
\unskip\
\newblock
\APACrefYearMonthDay{2003}{{\APACmonth{01}}}{}.
\newblock
{\BBOQ}\APACrefatitle {{Helicity patterns on the sun}} {{Helicity patterns on the sun}}.{\BBCQ}
\newblock
\APACjournalVolNumPages{Advances in Space Research}{32}{10}{1867-1874}.
\newblock
\begin{APACrefDOI} \doi{10.1016/S0273-1177(03)90620-X} \end{APACrefDOI}
\PrintBackRefs{\CurrentBib}

\bibitem [\protect \citeauthoryear {%
{Riley}%
}{%
{Riley}%
}{%
{\protect \APACyear {2012}}%
}]{%
riley2012frequency}
\APACinsertmetastar {%
riley2012frequency}%
\begin{APACrefauthors}%
{Riley}, P.%
\end{APACrefauthors}%
\unskip\
\newblock
\APACrefYearMonthDay{2012}{{\APACmonth{02}}}{}.
\newblock
{\BBOQ}\APACrefatitle {{On the probability of occurrence of extreme space weather events}} {{On the probability of occurrence of extreme space weather events}}.{\BBCQ}
\newblock
\APACjournalVolNumPages{Space Weather}{10}{2}{02012}.
\newblock
\begin{APACrefDOI} \doi{10.1029/2011SW000734} \end{APACrefDOI}
\PrintBackRefs{\CurrentBib}

\bibitem [\protect \citeauthoryear {%
{Ritter}%
\ \protect \BOthers {.}}{%
{Ritter}%
\ \protect \BOthers {.}}{%
{\protect \APACyear {2015}}%
}]{%
ritter2015}
\APACinsertmetastar {%
ritter2015}%
\begin{APACrefauthors}%
{Ritter}, B.%
, {Meskers}, A\BPBI J\BPBI H.%
, {Miles}, O.%
, {Ru{\ss}wurm}, M.%
, {Scully}, S.%
, {Rold{\'a}n}, A.%
\BDBL {}{Ruffenach}, A.%
\end{APACrefauthors}%
\unskip\
\newblock
\APACrefYearMonthDay{2015}{{\APACmonth{02}}}{}.
\newblock
{\BBOQ}\APACrefatitle {{A Space Weather Information Service Based Upon Remote and In-Situ Measurements of Coronal Mass Ejections Heading for Earth}} {{A Space Weather Information Service Based Upon Remote and In-Situ Measurements of Coronal Mass Ejections Heading for Earth}}.{\BBCQ}
\newblock
\APACjournalVolNumPages{Journ. Space Weather Space Climate}{5}{}{}.
\newblock
\begin{APACrefDOI} \doi{10.1051/swsc/2015006} \end{APACrefDOI}
\PrintBackRefs{\CurrentBib}

\bibitem [\protect \citeauthoryear {%
{Scherrer}%
\ \protect \BOthers {.}}{%
{Scherrer}%
\ \protect \BOthers {.}}{%
{\protect \APACyear {2012}}%
}]{%
scherrer2012HMI}
\APACinsertmetastar {%
scherrer2012HMI}%
\begin{APACrefauthors}%
{Scherrer}, P\BPBI H.%
, {Schou}, J.%
, {Bush}, R\BPBI I.%
, {Kosovichev}, A\BPBI G.%
, {Bogart}, R\BPBI S.%
, {Hoeksema}, J\BPBI T.%
\BDBL {}{Tomczyk}, S.%
\end{APACrefauthors}%
\unskip\
\newblock
\APACrefYearMonthDay{2012}{{\APACmonth{01}}}{}.
\newblock
{\BBOQ}\APACrefatitle {{The Helioseismic and Magnetic Imager (HMI) Investigation for the Solar Dynamics Observatory (SDO)}} {{The Helioseismic and Magnetic Imager (HMI) Investigation for the Solar Dynamics Observatory (SDO)}}.{\BBCQ}
\newblock
\APACjournalVolNumPages{\solphys}{275}{1-2}{207-227}.
\newblock
\begin{APACrefDOI} \doi{10.1007/s11207-011-9834-2} \end{APACrefDOI}
\PrintBackRefs{\CurrentBib}

\bibitem [\protect \citeauthoryear {%
{Schwartz}%
}{%
{Schwartz}%
}{%
{\protect \APACyear {1998}}%
}]{%
schwarz1998shock}
\APACinsertmetastar {%
schwarz1998shock}%
\begin{APACrefauthors}%
{Schwartz}, S\BPBI J.%
\end{APACrefauthors}%
\unskip\
\newblock
\APACrefYearMonthDay{1998}{{\APACmonth{01}}}{}.
\newblock
{\BBOQ}\APACrefatitle {{Shock and Discontinuity Normals, Mach Numbers, and Related Parameters}} {{Shock and Discontinuity Normals, Mach Numbers, and Related Parameters}}.{\BBCQ}
\newblock
\APACjournalVolNumPages{ISSI Scientific Reports Series}{1}{}{249-270}.
\PrintBackRefs{\CurrentBib}

\bibitem [\protect \citeauthoryear {%
Scolini%
\ \protect \BOthers {.}}{%
Scolini%
\ \protect \BOthers {.}}{%
{\protect \APACyear {2020}}%
}]{%
scolini2020}
\APACinsertmetastar {%
scolini2020}%
\begin{APACrefauthors}%
Scolini, C.%
, Chané, E.%
, Temmer, M.%
, Kilpua, E\BPBI K\BPBI J.%
, Dissauer, K.%
, Veronig, A\BPBI M.%
\BDBL {}Poedts, S.%
\end{APACrefauthors}%
\unskip\
\newblock
\APACrefYearMonthDay{2020}{feb}{}.
\newblock
{\BBOQ}\APACrefatitle {CME–CME Interactions as Sources of CME Geoeffectiveness: The Formation of the Complex Ejecta and Intense Geomagnetic Storm in 2017 Early September} {Cme–cme interactions as sources of cme geoeffectiveness: The formation of the complex ejecta and intense geomagnetic storm in 2017 early september}.{\BBCQ}
\newblock
\APACjournalVolNumPages{The Astrophysical Journal Supplement Series}{247}{1}{21}.
\newblock
\begin{APACrefURL} \url{https://dx.doi.org/10.3847/1538-4365/ab6216} \end{APACrefURL}
\newblock
\begin{APACrefDOI} \doi{10.3847/1538-4365/ab6216} \end{APACrefDOI}
\PrintBackRefs{\CurrentBib}

\bibitem [\protect \citeauthoryear {%
{Smith}%
\ \protect \BOthers {.}}{%
{Smith}%
\ \protect \BOthers {.}}{%
{\protect \APACyear {1998}}%
}]{%
smith1998ACEmag}
\APACinsertmetastar {%
smith1998ACEmag}%
\begin{APACrefauthors}%
{Smith}, C\BPBI W.%
, {L'Heureux}, J.%
, {Ness}, N\BPBI F.%
, {Acu{\~n}a}, M\BPBI H.%
, {Burlaga}, L\BPBI F.%
\BCBL {}\ \BBA {} {Scheifele}, J.%
\end{APACrefauthors}%
\unskip\
\newblock
\APACrefYearMonthDay{1998}{{\APACmonth{07}}}{}.
\newblock
{\BBOQ}\APACrefatitle {{The ACE Magnetic Fields Experiment}} {{The ACE Magnetic Fields Experiment}}.{\BBCQ}
\newblock
\APACjournalVolNumPages{\ssr}{86}{}{613-632}.
\newblock
\begin{APACrefDOI} \doi{10.1023/A:1005092216668} \end{APACrefDOI}
\PrintBackRefs{\CurrentBib}

\bibitem [\protect \citeauthoryear {%
Spogli%
\ \protect \BOthers {.}}{%
Spogli%
\ \protect \BOthers {.}}{%
{\protect \APACyear {2024}}%
}]{%
Spogli2024}
\APACinsertmetastar {%
Spogli2024}%
\begin{APACrefauthors}%
Spogli, L.%
, Alberti, T.%
, Bagiacchi, P.%
, cafarella, l.%
, Cesaroni, C.%
, Cianchini, G.%
\BDBL {}viola, m.%
\end{APACrefauthors}%
\unskip\
\newblock
\APACrefYearMonthDay{2024}{06}{}.
\newblock
{\BBOQ}\APACrefatitle {The effects of the May 2024 Mother's Day superstorm over the Mediterranean sector: from data to public communication} {The effects of the may 2024 mother's day superstorm over the mediterranean sector: from data to public communication}.{\BBCQ}
\newblock
\APACjournalVolNumPages{Annals of geophysics = Annali di geofisica}{67}{}{218}.
\newblock
\begin{APACrefDOI} \doi{10.4401/ag-9117} \end{APACrefDOI}
\PrintBackRefs{\CurrentBib}

\bibitem [\protect \citeauthoryear {%
{Stone}%
\ \protect \BOthers {.}}{%
{Stone}%
\ \protect \BOthers {.}}{%
{\protect \APACyear {1998}}%
}]{%
stone1998ace}
\APACinsertmetastar {%
stone1998ace}%
\begin{APACrefauthors}%
{Stone}, E\BPBI C.%
, {Frandsen}, A\BPBI M.%
, {Mewaldt}, R\BPBI A.%
, {Christian}, E\BPBI R.%
, {Margolies}, D.%
, {Ormes}, J\BPBI F.%
\BCBL {}\ \BBA {} {Snow}, F.%
\end{APACrefauthors}%
\unskip\
\newblock
\APACrefYearMonthDay{1998}{{\APACmonth{07}}}{}.
\newblock
{\BBOQ}\APACrefatitle {{The Advanced Composition Explorer}} {{The Advanced Composition Explorer}}.{\BBCQ}
\newblock
\APACjournalVolNumPages{\ssr}{86}{}{1-22}.
\newblock
\begin{APACrefDOI} \doi{10.1023/A:1005082526237} \end{APACrefDOI}
\PrintBackRefs{\CurrentBib}

\bibitem [\protect \citeauthoryear {%
Sugiura%
}{%
Sugiura%
}{%
{\protect \APACyear {1964}}%
}]{%
sugiura1964}
\APACinsertmetastar {%
sugiura1964}%
\begin{APACrefauthors}%
Sugiura, M.%
\end{APACrefauthors}%
\unskip\
\newblock
\APACrefYearMonthDay{1964}{}{}.
\newblock
{\BBOQ}\APACrefatitle {Hourly values of equatorial Dst for the IGY} {Hourly values of equatorial dst for the igy}.{\BBCQ}
\newblock
\APACjournalVolNumPages{Annals of the International Geophysical Year}{}{}{}.
\PrintBackRefs{\CurrentBib}

\bibitem [\protect \citeauthoryear {%
{SunPy Community}%
\ \protect \BOthers {.}}{%
{SunPy Community}%
\ \protect \BOthers {.}}{%
{\protect \APACyear {2020}}%
}]{%
sunpy_2020}
\APACinsertmetastar {%
sunpy_2020}%
\begin{APACrefauthors}%
{SunPy Community}%
, {Barnes}, W\BPBI T.%
, {Bobra}, M\BPBI G.%
, {Christe}, S\BPBI D.%
, {Freij}, N.%
, {Hayes}, L\BPBI A.%
\BDBL {}{Dang}, T\BPBI K.%
\end{APACrefauthors}%
\unskip\
\newblock
\APACrefYearMonthDay{2020}{{\APACmonth{02}}}{}.
\newblock
{\BBOQ}\APACrefatitle {{The SunPy Project: Open Source Development and Status of the Version 1.0 Core Package}} {{The SunPy Project: Open Source Development and Status of the Version 1.0 Core Package}}.{\BBCQ}
\newblock
\APACjournalVolNumPages{\apj}{890}{1}{68}.
\newblock
\begin{APACrefURL} \url{https://zenodo.org/records/13743565} \end{APACrefURL}
\newblock
\begin{APACrefDOI} \doi{10.3847/1538-4357/ab4f7a} \end{APACrefDOI}
\PrintBackRefs{\CurrentBib}

\bibitem [\protect \citeauthoryear {%
{Temerin}%
\ \BBA {} {Li}%
}{%
{Temerin}%
\ \BBA {} {Li}%
}{%
{\protect \APACyear {2006}}%
}]{%
Temerin2006}
\APACinsertmetastar {%
Temerin2006}%
\begin{APACrefauthors}%
{Temerin}, M.%
\BCBT {}\ \BBA {} {Li}, X.%
\end{APACrefauthors}%
\unskip\
\newblock
\APACrefYearMonthDay{2006}{{\APACmonth{04}}}{}.
\newblock
{\BBOQ}\APACrefatitle {{Dst model for 1995-2002}} {{Dst model for 1995-2002}}.{\BBCQ}
\newblock
\APACjournalVolNumPages{\jgr (Space Physics)}{111}{A4}{A04221}.
\newblock
\begin{APACrefDOI} \doi{10.1029/2005JA011257} \end{APACrefDOI}
\PrintBackRefs{\CurrentBib}

\bibitem [\protect \citeauthoryear {%
Temmer%
\ \protect \BOthers {.}}{%
Temmer%
\ \protect \BOthers {.}}{%
{\protect \APACyear {2023}}%
}]{%
temmer2023}
\APACinsertmetastar {%
temmer2023}%
\begin{APACrefauthors}%
Temmer, M.%
, Scolini, C.%
, Richardson, I\BPBI G.%
, Heinemann, S\BPBI G.%
, Paouris, E.%
, Vourlidas, A.%
\BDBL {}Zhuang, B.%
\end{APACrefauthors}%
\unskip\
\newblock
\APACrefYearMonthDay{2023}{}{}.
\newblock
{\BBOQ}\APACrefatitle {CME propagation through the heliosphere: Status and future of observations and model development} {Cme propagation through the heliosphere: Status and future of observations and model development}.{\BBCQ}
\newblock
\APACjournalVolNumPages{Advances in Space Research}{}{}{}.
\newblock
\begin{APACrefURL} \url{https://www.sciencedirect.com/science/article/pii/S0273117723005239} \end{APACrefURL}
\newblock
\begin{APACrefDOI} \doi{https://doi.org/10.1016/j.asr.2023.07.003} \end{APACrefDOI}
\PrintBackRefs{\CurrentBib}

\bibitem [\protect \citeauthoryear {%
Themens%
\ \protect \BOthers {.}}{%
Themens%
\ \protect \BOthers {.}}{%
{\protect \APACyear {2024}}%
}]{%
themens2024}
\APACinsertmetastar {%
themens2024}%
\begin{APACrefauthors}%
Themens, D\BPBI R.%
, Elvidge, S.%
, McCaffrey, A.%
, Jayachandran, P\BPBI T.%
, Coster, A.%
, Varney, R\BPBI H.%
\BDBL {}Reid, B.%
\end{APACrefauthors}%
\unskip\
\newblock
\APACrefYearMonthDay{2024}{}{}.
\newblock
{\BBOQ}\APACrefatitle {The High Latitude Ionospheric Response to the Major May 2024 Geomagnetic Storm: A Synoptic View} {The high latitude ionospheric response to the major may 2024 geomagnetic storm: A synoptic view}.{\BBCQ}
\newblock
\APACjournalVolNumPages{Geophysical Research Letters}{51}{19}{e2024GL111677}.
\newblock
\begin{APACrefURL} \url{https://agupubs.onlinelibrary.wiley.com/doi/abs/10.1029/2024GL111677} \end{APACrefURL}
\newblock
\APACrefnote{e2024GL111677 2024GL111677}
\newblock
\begin{APACrefDOI} \doi{https://doi.org/10.1029/2024GL111677} \end{APACrefDOI}
\PrintBackRefs{\CurrentBib}

\bibitem [\protect \citeauthoryear {%
{Vourlidas}%
, {Patsourakos}%
\BCBL {}\ \BBA {} {Savani}%
}{%
{Vourlidas}%
\ \protect \BOthers {.}}{%
{\protect \APACyear {2019}}%
}]{%
vourlidas2019review}
\APACinsertmetastar {%
vourlidas2019review}%
\begin{APACrefauthors}%
{Vourlidas}, A.%
, {Patsourakos}, S.%
\BCBL {}\ \BBA {} {Savani}, N\BPBI P.%
\end{APACrefauthors}%
\unskip\
\newblock
\APACrefYearMonthDay{2019}{{\APACmonth{07}}}{}.
\newblock
{\BBOQ}\APACrefatitle {{Predicting the geoeffective properties of coronal mass ejections: current status, open issues and path forward}} {{Predicting the geoeffective properties of coronal mass ejections: current status, open issues and path forward}}.{\BBCQ}
\newblock
\APACjournalVolNumPages{Philosophical Transactions of the Royal Society of London Series A}{377}{2148}{20180096}.
\newblock
\begin{APACrefDOI} \doi{10.1098/rsta.2018.0096} \end{APACrefDOI}
\PrintBackRefs{\CurrentBib}

\bibitem [\protect \citeauthoryear {%
{Vr{\v{s}}nak}%
\ \protect \BOthers {.}}{%
{Vr{\v{s}}nak}%
\ \protect \BOthers {.}}{%
{\protect \APACyear {2013}}%
}]{%
vrsnak2013propagation}
\APACinsertmetastar {%
vrsnak2013propagation}%
\begin{APACrefauthors}%
{Vr{\v{s}}nak}, B.%
, {{\v{Z}}ic}, T.%
, {Vrbanec}, D.%
, {Temmer}, M.%
, {Rollett}, T.%
, {M{\"o}stl}, C.%
\BDBL {}{Shanmugaraju}, A.%
\end{APACrefauthors}%
\unskip\
\newblock
\APACrefYearMonthDay{2013}{Jul}{}.
\newblock
{\BBOQ}\APACrefatitle {{Propagation of Interplanetary Coronal Mass Ejections: The Drag-Based Model}} {{Propagation of Interplanetary Coronal Mass Ejections: The Drag-Based Model}}.{\BBCQ}
\newblock
\APACjournalVolNumPages{\solphys}{285}{1-2}{295-315}.
\newblock
\begin{APACrefDOI} \doi{10.1007/s11207-012-0035-4} \end{APACrefDOI}
\PrintBackRefs{\CurrentBib}

\bibitem [\protect \citeauthoryear {%
Wang%
, Liu%
, Zhao%
\BCBL {}\ \BBA {} Hu%
}{%
Wang%
\ \protect \BOthers {.}}{%
{\protect \APACyear {2024}}%
}]{%
wang2024}
\APACinsertmetastar {%
wang2024}%
\begin{APACrefauthors}%
Wang, R.%
, Liu, Y\BPBI D.%
, Zhao, X.%
\BCBL {}\ \BBA {} Hu, H.%
\end{APACrefauthors}%
\unskip\
\newblock
\APACrefYearMonthDay{2024}{}{}.
\newblock
\APACrefbtitle {Unveiling Key Factors in the Solar Eruptions Leading to the Solar Superstorm in 2024 May.} {Unveiling key factors in the solar eruptions leading to the solar superstorm in 2024 may.}
\newblock
\begin{APACrefURL} \url{https://arxiv.org/abs/2410.00891} \end{APACrefURL}
\PrintBackRefs{\CurrentBib}

\bibitem [\protect \citeauthoryear {%
Wanliss%
\ \BBA {} Showalter%
}{%
Wanliss%
\ \BBA {} Showalter%
}{%
{\protect \APACyear {2006}}%
}]{%
wanliss2006}
\APACinsertmetastar {%
wanliss2006}%
\begin{APACrefauthors}%
Wanliss, J\BPBI A.%
\BCBT {}\ \BBA {} Showalter, K\BPBI M.%
\end{APACrefauthors}%
\unskip\
\newblock
\APACrefYearMonthDay{2006}{}{}.
\newblock
{\BBOQ}\APACrefatitle {High-resolution global storm index: Dst versus SYM-H} {High-resolution global storm index: Dst versus sym-h}.{\BBCQ}
\newblock
\APACjournalVolNumPages{Journal of Geophysical Research: Space Physics}{111}{A2}{}.
\newblock
\begin{APACrefURL} \url{https://agupubs.onlinelibrary.wiley.com/doi/abs/10.1029/2005JA011034} \end{APACrefURL}
\newblock
\begin{APACrefDOI} \doi{https://doi.org/10.1029/2005JA011034} \end{APACrefDOI}
\PrintBackRefs{\CurrentBib}

\bibitem [\protect \citeauthoryear {%
Weiler%
\ \protect \BOthers {.}}{%
Weiler%
\ \protect \BOthers {.}}{%
{\protect \APACyear {2024}}%
}]{%
Weiler2024}
\APACinsertmetastar {%
Weiler2024}%
\begin{APACrefauthors}%
Weiler, E.%
, Davies, E.%
, M{\"o}stl, C.%
, Amerstorfer, T.%
, {Le Lou{\"e}dec}, J.%
\BCBL {}\ \BBA {} Bauer, M.%
\end{APACrefauthors}%
\unskip\
\newblock
\APACrefYearMonthDay{2024}{}{}.
\newblock
\APACrefbtitle {{May 2024 Superstorm}.} {{May 2024 Superstorm}.}
\newblock
\APACaddressPublisher{}{figshare}.
\newblock
\begin{APACrefURL} \url{https://figshare.com/articles/dataset/May_2024_Superstorm/27792873} \end{APACrefURL}
\newblock
\begin{APACrefDOI} \doi{10.6084/m9.figshare.27792873.v2} \end{APACrefDOI}
\PrintBackRefs{\CurrentBib}

\bibitem [\protect \citeauthoryear {%
Weiler%
\ \protect \BOthers {.}}{%
Weiler%
\ \protect \BOthers {.}}{%
{\protect \APACyear {2025}}%
}]{%
weiler_2025_14772679}
\APACinsertmetastar {%
weiler_2025_14772679}%
\begin{APACrefauthors}%
Weiler, E.%
, Möstl, C.%
, Davies, E.%
, Veronig, A.%
, Amerstorfer, U.%
, Amerstorfer, T.%
\BDBL {}Reiss, M.%
\end{APACrefauthors}%
\unskip\
\newblock
\APACrefYearMonthDay{2025}{{\APACmonth{01}}}{}.
\newblock
\APACrefbtitle {Software for "First observations of a geomagnetic superstorm with a sub-L1 monitor".} {Software for "first observations of a geomagnetic superstorm with a sub-l1 monitor".}
\newblock
\APACaddressPublisher{}{Zenodo}.
\newblock
\begin{APACrefURL} \url{https://doi.org/10.5281/zenodo.14772679} \end{APACrefURL}
\newblock
\begin{APACrefDOI} \doi{10.5281/zenodo.14772679} \end{APACrefDOI}
\PrintBackRefs{\CurrentBib}

\bibitem [\protect \citeauthoryear {%
{Weiss}%
, {M{\"o}stl}%
, {Amerstorfer}%
\BCBL {}\ \protect \BOthers {.}}{%
{Weiss}%
, {M{\"o}stl}%
, {Amerstorfer}%
\BCBL {}\ \protect \BOthers {.}}{%
{\protect \APACyear {2021}}%
}]{%
weiss2021analysis}
\APACinsertmetastar {%
weiss2021analysis}%
\begin{APACrefauthors}%
{Weiss}, A\BPBI J.%
, {M{\"o}stl}, C.%
, {Amerstorfer}, T.%
, {Bailey}, R\BPBI L.%
, {Reiss}, M\BPBI A.%
, {Hinterreiter}, J.%
\BDBL {}{Bauer}, M.%
\end{APACrefauthors}%
\unskip\
\newblock
\APACrefYearMonthDay{2021}{{\APACmonth{01}}}{}.
\newblock
{\BBOQ}\APACrefatitle {{Analysis of Coronal Mass Ejection Flux Rope Signatures Using 3DCORE and Approximate Bayesian Computation}} {{Analysis of Coronal Mass Ejection Flux Rope Signatures Using 3DCORE and Approximate Bayesian Computation}}.{\BBCQ}
\newblock
\APACjournalVolNumPages{\apjs}{252}{1}{9}.
\newblock
\begin{APACrefDOI} \doi{10.3847/1538-4365/abc9bd} \end{APACrefDOI}
\PrintBackRefs{\CurrentBib}

\bibitem [\protect \citeauthoryear {%
{Weiss}%
, {M{\"o}stl}%
, {Davies}%
\BCBL {}\ \protect \BOthers {.}}{%
{Weiss}%
, {M{\"o}stl}%
, {Davies}%
\BCBL {}\ \protect \BOthers {.}}{%
{\protect \APACyear {2021}}%
}]{%
weiss2021triple}
\APACinsertmetastar {%
weiss2021triple}%
\begin{APACrefauthors}%
{Weiss}, A\BPBI J.%
, {M{\"o}stl}, C.%
, {Davies}, E\BPBI E.%
, {Amerstorfer}, T.%
, {Bauer}, M.%
, {Hinterreiter}, J.%
\BDBL {}{Baumjohann}, W.%
\end{APACrefauthors}%
\unskip\
\newblock
\APACrefYearMonthDay{2021}{{\APACmonth{12}}}{}.
\newblock
{\BBOQ}\APACrefatitle {{Multi-point analysis of coronal mass ejection flux ropes using combined data from Solar Orbiter, BepiColombo, and Wind}} {{Multi-point analysis of coronal mass ejection flux ropes using combined data from Solar Orbiter, BepiColombo, and Wind}}.{\BBCQ}
\newblock
\APACjournalVolNumPages{\aap}{656}{}{A13}.
\newblock
\begin{APACrefDOI} \doi{10.1051/0004-6361/202140919} \end{APACrefDOI}
\PrintBackRefs{\CurrentBib}

\bibitem [\protect \citeauthoryear {%
{Zhang}%
\ \protect \BOthers {.}}{%
{Zhang}%
\ \protect \BOthers {.}}{%
{\protect \APACyear {2007}}%
}]{%
zhang2007}
\APACinsertmetastar {%
zhang2007}%
\begin{APACrefauthors}%
{Zhang}, J.%
, {Richardson}, I\BPBI G.%
, {Webb}, D\BPBI F.%
, {Gopalswamy}, N.%
, {Huttunen}, E.%
, {Kasper}, J\BPBI C.%
\BDBL {}{Zhukov}, A\BPBI N.%
\end{APACrefauthors}%
\unskip\
\newblock
\APACrefYearMonthDay{2007}{{\APACmonth{10}}}{}.
\newblock
{\BBOQ}\APACrefatitle {{Solar and interplanetary sources of major geomagnetic storms (Dst <= -100 nT) during 1996-2005}} {{Solar and interplanetary sources of major geomagnetic storms (Dst <= -100 nT) during 1996-2005}}.{\BBCQ}
\newblock
\APACjournalVolNumPages{Journal of Geophysical Research (Space Physics)}{112}{A10}{A10102}.
\newblock
\begin{APACrefDOI} \doi{10.1029/2007JA012321} \end{APACrefDOI}
\PrintBackRefs{\CurrentBib}

\bibitem [\protect \citeauthoryear {%
Čalogović%
\ \protect \BOthers {.}}{%
Čalogović%
\ \protect \BOthers {.}}{%
{\protect \APACyear {2021}}%
}]{%
calogovic2021dbem}
\APACinsertmetastar {%
calogovic2021dbem}%
\begin{APACrefauthors}%
Čalogović, J.%
, Dumbović, M.%
, Sudar, D.%
, Vršnak, B.%
, Martinić, K.%
, Temmer, M.%
\BCBL {}\ \BBA {} Veronig, A\BPBI M.%
\end{APACrefauthors}%
\unskip\
\newblock
\APACrefYearMonthDay{2021}{{\APACmonth{07}}}{}.
\newblock
{\BBOQ}\APACrefatitle {Probabilistic Drag-Based Ensemble Model (DBEM) Evaluation for Heliospheric Propagation of CMEs} {Probabilistic drag-based ensemble model (dbem) evaluation for heliospheric propagation of cmes}.{\BBCQ}
\newblock
\APACjournalVolNumPages{Solar Physics}{296}{7}{}.
\newblock
\begin{APACrefURL} \url{http://dx.doi.org/10.1007/s11207-021-01859-5} \end{APACrefURL}
\newblock
\begin{APACrefDOI} \doi{10.1007/s11207-021-01859-5} \end{APACrefDOI}
\PrintBackRefs{\CurrentBib}

\end{thebibliography}
%

\end{document}